\documentclass[letterpaper,twocolumn,10pt]{article}
\usepackage{usenix2019_v3}

\usepackage{tikz}
\usepackage{amsmath}
\usepackage{enumitem}
\usepackage{subcaption}
\usepackage{graphicx}
\usepackage{booktabs}
\usepackage{makecell}
\usepackage{amsmath}
\usepackage{mathtools}
\usepackage{breqn}
\usepackage{multirow}

\usepackage{soul}

\usepackage{pifont}

\usepackage[available,functional]{usenixbadges}

\definecolor{ForestGreen}{HTML}{009B55}
\definecolor{OliveGreen}{HTML}{3C8031}

\newcommand{\cmark}{\textcolor{green}{\ding{51}}}
\newcommand{\xmark}{\textcolor{red}{\ding{55}}}

\newcommand{\TODO}[1]{\textcolor{red}{#1}} 
\newcommand{\OD}[1]{\textcolor{cyan}{#1}} 
\newcommand{\jb}[1]{\textcolor{magenta}{#1}} 

\begin{document}

\date{}
\pagestyle{empty} 
\title{\Large \bf Preparation Meets Opportunity:\\Enhancing Data Preprocessing for ML Training With Seneca}

\author{
{\rm Omkar Desai} \\
Syracuse University
\and
{\rm Ziyang Jiao} \\
Syracuse University
\and
{\rm Shuyi Pei} \\
Samsung Semiconductor
\and
{\rm Janki Bhimani} \\
Florida International University
\and
{\rm Bryan S. Kim}\\
Syracuse University
}
\maketitle

\begin{abstract}
Input data preprocessing is a common bottleneck when concurrently training multimedia machine learning (ML) models in modern systems. 
To alleviate these bottlenecks and reduce the training time for concurrent jobs, 
we present Seneca, a data loading system that optimizes cache partitioning and data sampling for the data storage and ingestion (DSI) pipeline.
The design of Seneca contains two key techniques.
First, Seneca uses a performance model for the data pipeline to optimally partition the cache for three different forms of data (encoded, decoded, and augmented).
Second, Seneca opportunistically serves cached data over uncached ones during random batch sampling so that concurrent jobs benefit from each other.
We implement Seneca by modifying PyTorch and demonstrate its effectiveness by comparing it against several state-of-the-art caching systems for DNN training. 
Seneca reduces the makespan by 45.23\% compared to PyTorch and increases data processing throughput by up to 3.45$\times$
compared to the next best dataloader.

\end{abstract}

\section{Introduction}
Input data preprocessing is an essential step in all machine learning (ML) training jobs. During this, the data storage and ingestion (DSI) 
pipeline fetches samples from storage, decodes them into tensors, transforms and augments them as required, and loads them into the GPU for training.
These tasks are resource-intensive, requiring high I/O bandwidth and compute parallelism. 
The throughput of the DSI pipeline has a significant impact on training performance as GPUs depend on data from the DSI pipeline to train models.

\begin{figure}[t!]
    \begin{subfigure}[t!]{0.49\columnwidth}
        \includegraphics[width=\linewidth]{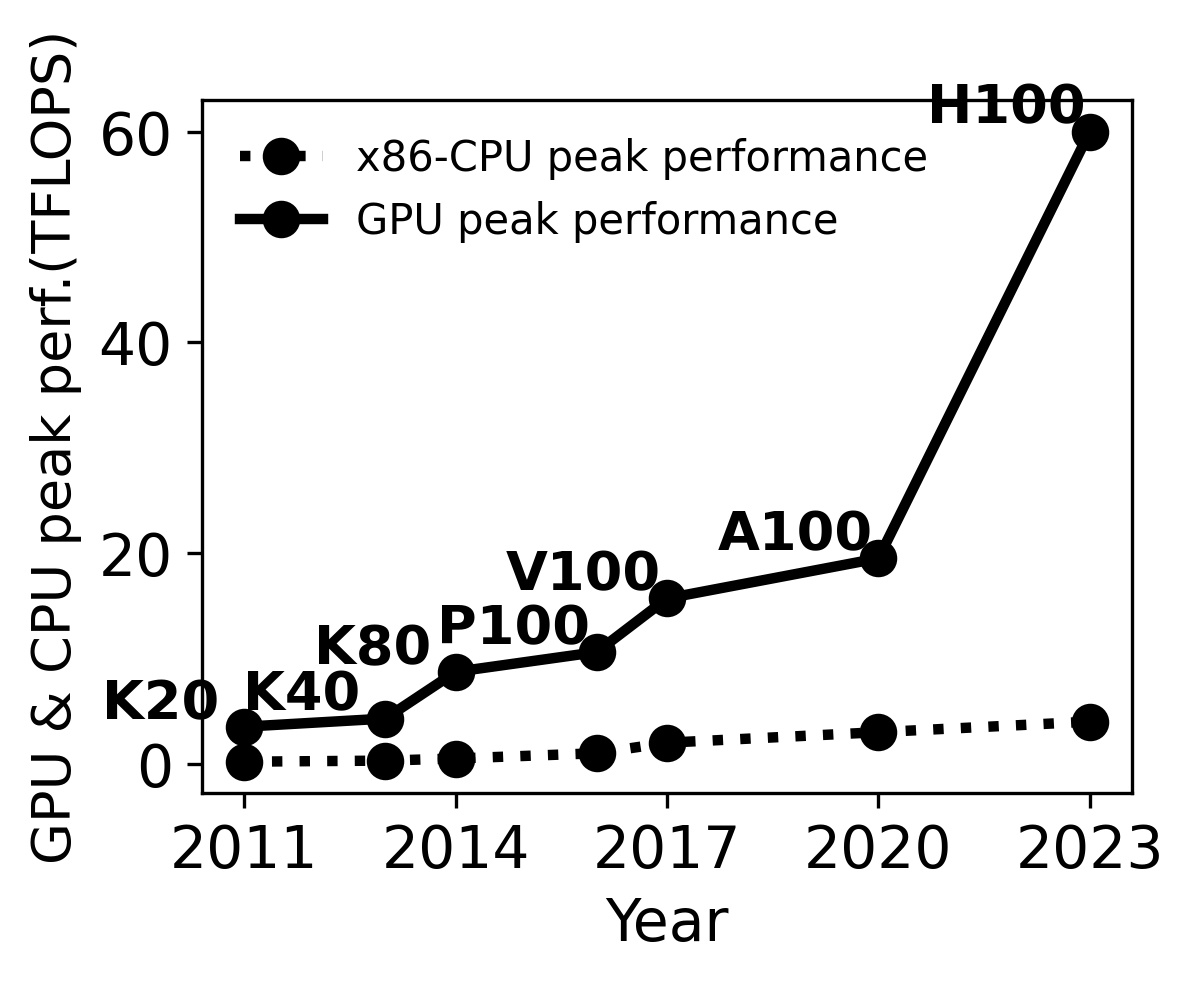}
        \caption{
        Growing gap between CPU and GPU TFLOPS from 2011 to 2023.
        }
        \label{fig:gpu_v_cpu_flops}
    \end{subfigure}%
    \hfill
    \begin{subfigure}[t!]{0.49\columnwidth}
        \includegraphics[width=\linewidth]{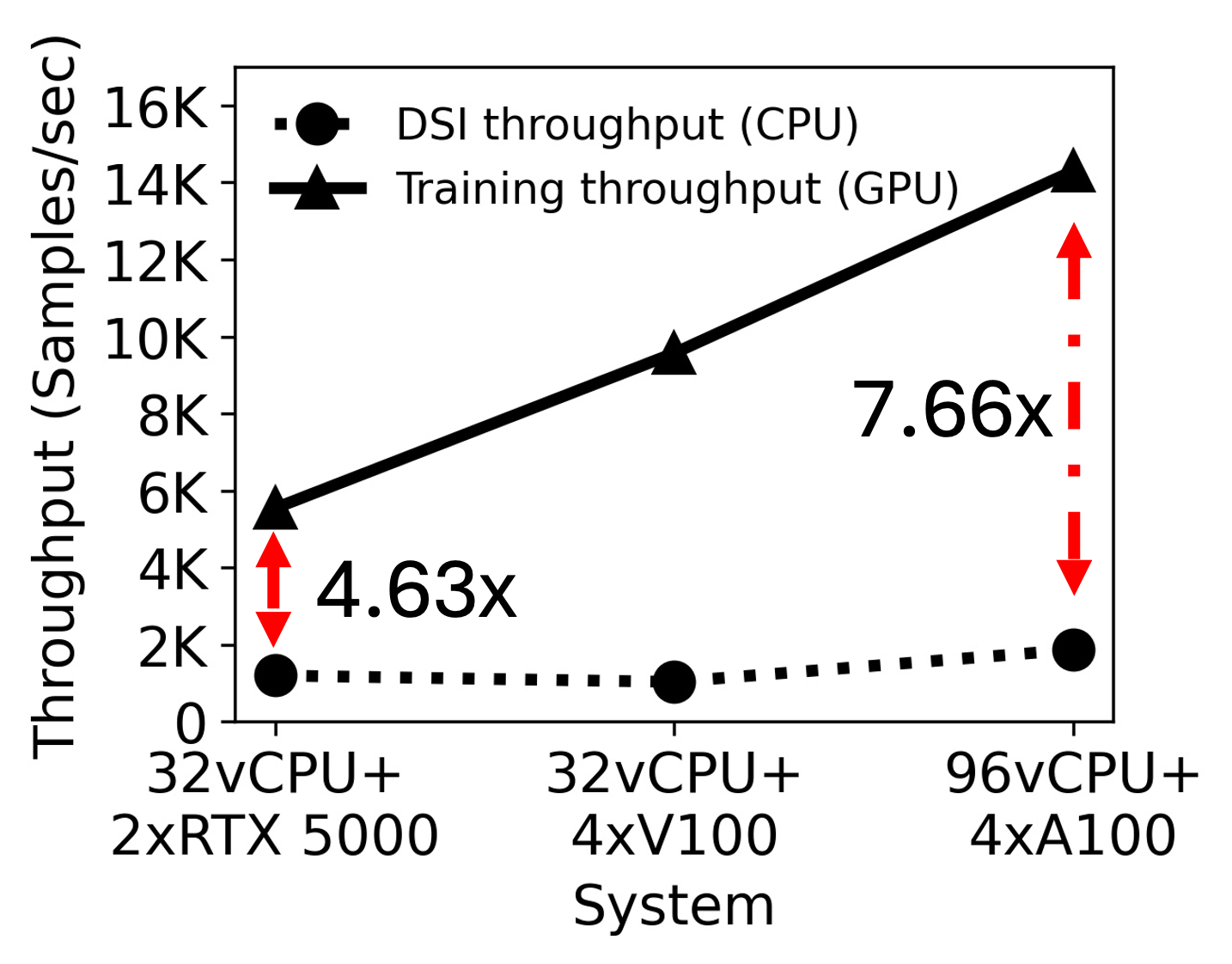}
        \caption{
        Multimedia model DSI vs GPU training throughput on various systems.
        }
        \label{fig:gpu_v_cpu_flops_throughput}
    \end{subfigure}
    \caption{
    The growing gap between CPU and GPU peak performance in TFLOPS from 2011--2023~\cite{awsp2instancetype,awsp3instancetype,AzureNCv4,gcp, nvidiak20, nvidiak40,nvidiak80,nvidiap100,nvidiav100,nvidiaa100,nvidiah100} (Figure~\ref{fig:gpu_v_cpu_flops}) resulting in DSI pipeline bottlenecks for multimedia model training (Figure~\ref{fig:gpu_v_cpu_flops_throughput}). 
    }
    
    \label{fig:gpu_v_cpu}
\end{figure}

With ML jobs training concurrently to maximize GPU utilization, the DSI pipeline has become a bottleneck, especially for image, video, audio, and recommendation models~\cite{murray2021tf, graur2024pecan, zhao2022understanding, lim2021zico, tensorflow2015-whitepaper, pytorch}.
Figure~\ref{fig:gpu_v_cpu} illustrates hardware trends affecting ML training. Figure~\ref{fig:gpu_v_cpu_flops} shows the widening gap between popular NVIDIA GPUs used for training and CPU performance. 
Figure~\ref{fig:gpu_v_cpu_flops_throughput} illustrates the impact of the CPU-GPU performance gap on multimedia model training by comparing the DSI throughput without training and the training throughput without DSI for SwinT~\cite{liu2021swin}, a transformer model on three real systems. The dotted line shows the upper bound of DSI throughput, while the solid line reflects the upper bound of training throughput, revealing that DSI is the bottleneck.
The disparity between DSI and training throughput increases from 4.63$\times$ on the RTX 5000 server to 7.66$\times$ on the A100 server, 
indicating that the DSI pipeline bottleneck is exacerbating.
The combination of a relatively slow CPU limits preprocessing performance, and the limited host DRAM forces more frequent data retrieval from slow storage, leading to training stalls and an overall performance degradation~\cite{um2023fastflow, kuchnik2022plumber, weng2022mlaas}.

While several prior works propose caching as a solution for improving the DSI pipeline throughput~\cite{graur2022cachew,lee2021refurbish,kumar2020quiver,mohan2020analyzing,khan2023shade,graur2024pecan}, 
we identify two unique challenges that limit its effectiveness. 
First, multiple intermediate forms of data exist in the pipeline, and which form to cache is not straightforward and remains unexplored. 
Data can exist in one of three forms: encoded for high storage density, decoded into tensors, and augmented randomly for training.
Determining which form to cache is subtle because 
encoded data requires more CPU-driven preprocessing~\cite{graur2024pecan,um2023fastflow}, 
while data in later stages (e.g., augmented) take up more space (up to 15$\times$~\cite{isenko2022my}). 
The optimal form to cache depends on training job parameters, cache capacity, and the performance of the slowest DSI component, factors not considered by previous works.

Second, current cache designs and data sampling for the DSI pipeline perform poorly when multiple jobs train models concurrently over the same dataset.
More specifically, the inherent nature of random sampling makes poor use of cache as data are sampled agnostic of what is available in the cache. 
For example, SHADE~\cite{khan2023shade} implements an importance-based sampling but its approach is incompatible with concurrent training, 
and Quiver~\cite{kumar2020quiver} presents a substitution-based sampling compatible with concurrent training but suffers from high bandwidth contention due to over-sampling.

We address these challenges by designing the following:
\begin{enumerate}[leftmargin=*, itemsep=-0.5ex] 
    \item \textbf{Model-Driven Partitioning (MDP).} 
    We build a high-level performance model for the DSI pipeline based on the insight that we can estimate the cache hit rate due to random sampling. This model is then used to predict the data pipeline throughput, constrained by the slowest component, and allows Seneca to partition the cache optimally for different data forms. 
    \item \textbf{Opportunistic Data Sampling (ODS).} 
    Based on the insight that a training job may benefit from the other's data sampling activities that share the same dataset, we design an opportunistic data sampler that serves cached data over uncached ones. This is possible because the sampling sequence does not need to adhere to the predetermined pseudo-random sequence as long as it appears random and each training data is consumed once in an epoch. 
\end{enumerate}
Preparation meets opportunity in Seneca by combining the design principles of MDP and ODS: integrating cache partitioning and data sampling to efficiently alleviate preprocessing bottlenecks in ML training.  
We implement Seneca by modifying the PyTorch~\cite{pytorch} dataloader and using Redis~\cite{redis} to cache data. 
Seneca improves DSI pipeline throughput by up to 3.45$\times$
compared to the next best dataloader and reduces makespan by 45.23\% over PyTorch, without compromising training accuracy.

The contributions of this work are as follows:
\begin{itemize}[leftmargin=*, itemsep=-0.5pt]
\item We design a performance model for the DSI pipeline in distributed ML training and develop a model-driven partitioning (MDP) scheme that uses the performance model to optimize cache partition sizes and improve training throughput. 
MDP is applicable to all types of ML training jobs but benefits preprocessing heavy jobs the most.
\item We develop an opportunistic data sampling (ODS) scheme 
that improves the cache hit rate and multi-job training throughput by replacing uncached data with cached ones while maintaining a pseudo-random order. The effectiveness of ODS increases with more concurrently running jobs.
\item We demonstrate the effectiveness of Seneca by evaluating it against five state-of-the-art solutions across seven models (3.4--633.4 million parameters) and three datasets (142GB--1.4TB) on five hardware configurations.  
\item The artifacts of Seneca are open source and available at ~\url{https://github.com/swiftomkar/seneca-fast26-pytorch}

\end{itemize}
\section{Background}

\begin{figure}
    \centering
    \includegraphics[width=\columnwidth]{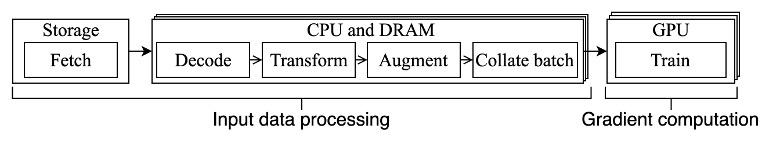}
    \vspace{-7pt}
    \caption{
        The ML training process: (1) fetches data from storage, (2) decodes, transforms, and forms a minibatch, and (3) loads the minibatch into the GPU for training. This repeats until the target accuracy is reached. Encoded data is small, but decoded and augmented data are much larger.
    }
    \label{fig:dsi_transformation_flow}
\end{figure}

\begin{table*}
\centering
\footnotesize
\caption{DSI pipeline for various ML model types. All types of training data undergo preprocessing before training. For all model types, decoding converts a dataset file into a tensor, while transformation and augmentation steps apply operations on the decoded tensor before collating samples into a batch.}
\resizebox{\textwidth}{!}{%
\begin{tabular}{l c c c c c}
\hline
\multirow{2}{*}{Model type} & \multicolumn{4}{c}{Preprocessing steps} & \multirow{2}{*}{Resource demand} \\
\cline{2-5}
 & Decode & Transform & Augment & Collate & \\ \hline
Image & Image file $\rightarrow$ Tensor & \makecell[cc]{Resize \\ Normalize} & \makecell[cc]{Random crop \\ Random flip} & [images:labels] & High \\ 
\hline
Audio & Audio file $\rightarrow$ Tensor & \makecell[cc]{Fourier transform \\ Padding} & \makecell[cc]{Time stretch \\ Time masking} & [audio:labels] & High \\
\hline
Text & Text file $\rightarrow$ Tensor & \makecell[cc]{Padding \\ Truncation} & \makecell[cc]{Shuffling \\ Masking} & [text:labels] & Low \\
\hline
Recommendation & Tabular $\rightarrow$ Tensor & \makecell[cc]{Padding \\ Truncation} & \makecell[cc]{Shuffling \\ Masking} & \makecell[cc]{[feature vector:\\user action]} & High \\
\hline
\end{tabular}%
}
\label{tab:dsi_pipelines}
\end{table*}

ML models are trained over multiple rounds called epochs. During an epoch, the entire dataset is processed exactly once. Each epoch is divided into multiple iterations, where each iteration processes a random, non-overlapping subset of data known as a minibatch. Training continues until further epochs no longer improve the model's accuracy.


Figure~\ref{fig:dsi_transformation_flow} outlines the two main components of the ML training process: (1) input data preprocessing and (2) gradient computation. 
In preprocessing, data is fetched from storage, decoded, transformed, augmented, and collated into mini-batches before being loaded onto the GPU for gradient computation. 
While decoding and collation are static preprocessing operations, transformations and augmentations are randomly applied to improve model generality.
Table~\ref{tab:dsi_pipelines} describes the data preprocessing steps for various ML model types and their resource demands. We note that multimedia and high-dimensional data have resource-intensive DSI pipelines due to larger sample sizes. 
During gradient computation, the model updates its weights and biases based on the training data.

ML model training is resource-intensive and time-consuming, so large GPU clusters are often used to speed up training through techniques like data parallelism~\cite{raina2009large}, model parallelism~\cite{dean2012large}, and pipelined parallelism~\cite{huang2019gpipe}. Our work focuses on optimizing the data preprocessing bottleneck commonly seen in multimedia ML (image, audio, and video) and deep learning recommendation models (DLRM) that use data parallelism~\cite{mohan2022looking, zhao2022understanding}. In data parallelism, the dataset is divided and processed in parallel across multiple GPUs, each running a copy of the same model. After every batch, the models are synchronized across GPUs using centralized methods (e.g., parameter servers~\cite{li2014communication}) or decentralized methods (e.g., all-reduce~\cite{bao2020preemptive} or ring-reduce~\cite{tang2020communication}). 

The following are the three main steps in the DSI pipeline. 
\begin{enumerate}[leftmargin=*, itemsep=-0.5pt]
\item \textbf{Fetching data from storage.} The first step in ML data preprocessing is reading data from the slow, remote storage service. DSI pipelines manage datasets from a few gigabytes to several petabytes in low-cost storage. To avoid the memorization of specific patterns, data samples must be fetched in random order, making caching difficult~\cite{lecun2002efficient, khan2023shade}.


\item \textbf{Transforming data.} Before training, each data sample must undergo several processing steps (illustrated in Table~\ref{tab:dsi_pipelines}) which are typically applied on the fly~\cite{graur2024pecan}. These transformations performed by the DSI pipeline can be computationally intensive and generally run on the CPU to support user-defined transformations~\cite{murray2021tf}. Common transformations necessary for all training jobs include decompressing, decoding, and collating samples. Additionally, every job also has specific static transformations (e.g., tokenization for text, quantization for audio) and random augmentations (e.g., image rotation, audio stretching) to improve model accuracy and generality~\cite{murray2021tf, AudibertCGKST23socc}. Random augmentations are especially critical to training generalizable models for image classification, object detection, and speech recognition. 


\item \textbf{Loading data to the GPU.} The final step is loading data to GPUs for gradient computation. To prevent stalls, the DSI pipeline must provide data at or above the GPU's data ingestion rate, which depends on the model size, training algorithm's arithmetic intensity, and the GPU's FLOPS.
\end{enumerate}

\section{Existing approaches}
We categorize works related to Seneca into four categories: system modeling, preprocessing optimization, hardware acceleration, and cache optimizations. 

\noindent \textbf{System modeling:} 
Lobster~\cite{liu2022lobster} models the I/O and DSI pipeline of a distributed training system to balance I/O and preprocessing threads.
NoPFS~\cite{dryden2021clairvoyant} builds a performance model for a distributed storage system given a clairvoyant knowledge of data sampling, and uses this to prefetch data from slower storage to cache.
Our work is inspired by high-level system modeling and builds a mathematical model of the entire DSI pipeline. 
Our model predicts the throughput for different forms of data in the cache and can be used to partition cache space and minimize preprocessing bottlenecks.

\noindent \textbf{Preprocessing optimization:} 
Several prior works have identified preprocessing of training data as a major bottleneck. Cedar~\cite{zhao2024cedar}, FastFlow~\cite{um2023fastflow}, GoldMiner~\cite{zhao2023goldminer}, Cachew~\cite{graur2022cachew}, and Pecan~\cite{graur2024pecan} address this issue by optimizing and offloading preprocessing tasks to remote CPUs based on profiled system metrics. Our work cost-effectively optimizes the DSI pipeline, without relying on remote CPUs.
PRESTO~\cite{isenko2022my} and Revamper~\cite{lee2021refurbish} explore caching partially or fully processed data to minimize repeated preprocessing computations on the same data, but don't consider their data inflation in conjunction with resource constraints presented by the training hardware.
Our work uses this insight and allocates an optimal ratio of the available cache to each form.


\noindent \textbf{Hardware acceleration:} 
DALI~\cite{nvidia-dali}, TrainBox~\cite{park2020trainbox}, and FusionFlow~\cite{kim2023fusionflow} uses hardware accelerators (GPUs and FPGAs) for preprocessing. This makes it challenging to apply user-defined transforms while also making poor use of expensive accelerators that are optimized for SIMD operations that differ from the stochastic nature of random augmentations. While our implementation is based on using CPUs for preprocessing, the core concepts of our work can be extended to data loaders that leverage hardware accelerators.

\noindent \textbf{Cache optimization:} 
Due to I/O bottlenecks and large dataset sizes in DNN training, many works focus on reducing data fetch time through caching. MINIO~\cite{mohan2020analyzing} uses a shared cache with a no-eviction policy to avoid thrashing, but its cache hit rate is limited by the cache-to-dataset size ratio. SHADE~\cite{khan2023shade} and iCache~\cite{chen2023icache} use importance sampling~\cite{katharopoulos2018not}, but since sample importance varies across jobs, sharing a cache for concurrent jobs is difficult. Quiver~\cite{kumar2020quiver} allows sample substitution from the cache but suffers from high oversampling overhead. Our approach introduces a cache-aware opportunistic sampler that prioritizes cached samples over uncached ones while maintaining the randomness and uniqueness needed for high model accuracy.


\section{Challenges for the DSI pipeline}
Based on the required data preprocessing and random data sampling of DNN model training, 
We describe below the two key challenges in effectively using limited cache space to alleviate the DSI pipeline bottleneck.

\subsection{Determining which data form to cache}\label{sec:motivation_1}
\begin{figure}[t!]
        \begin{subfigure}[t!]{0.49\columnwidth}
                \includegraphics[width=\linewidth]{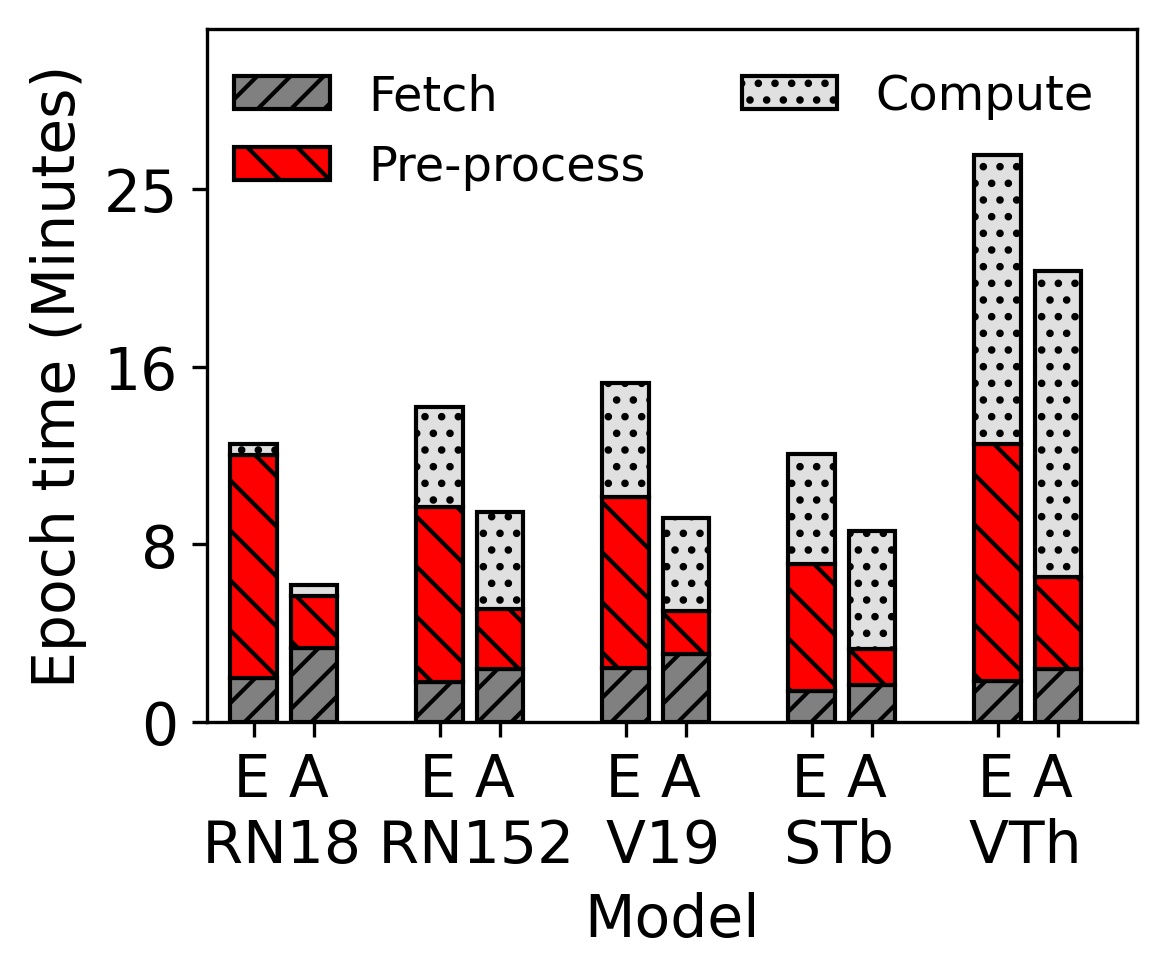}
                \caption{Epoch completion times with 450GB cache}
                \label{fig:fetch_prep_compute_450g}
        \end{subfigure}%
        \hfill
        \begin{subfigure}[t!]{0.49\columnwidth}
                \includegraphics[width=\linewidth]{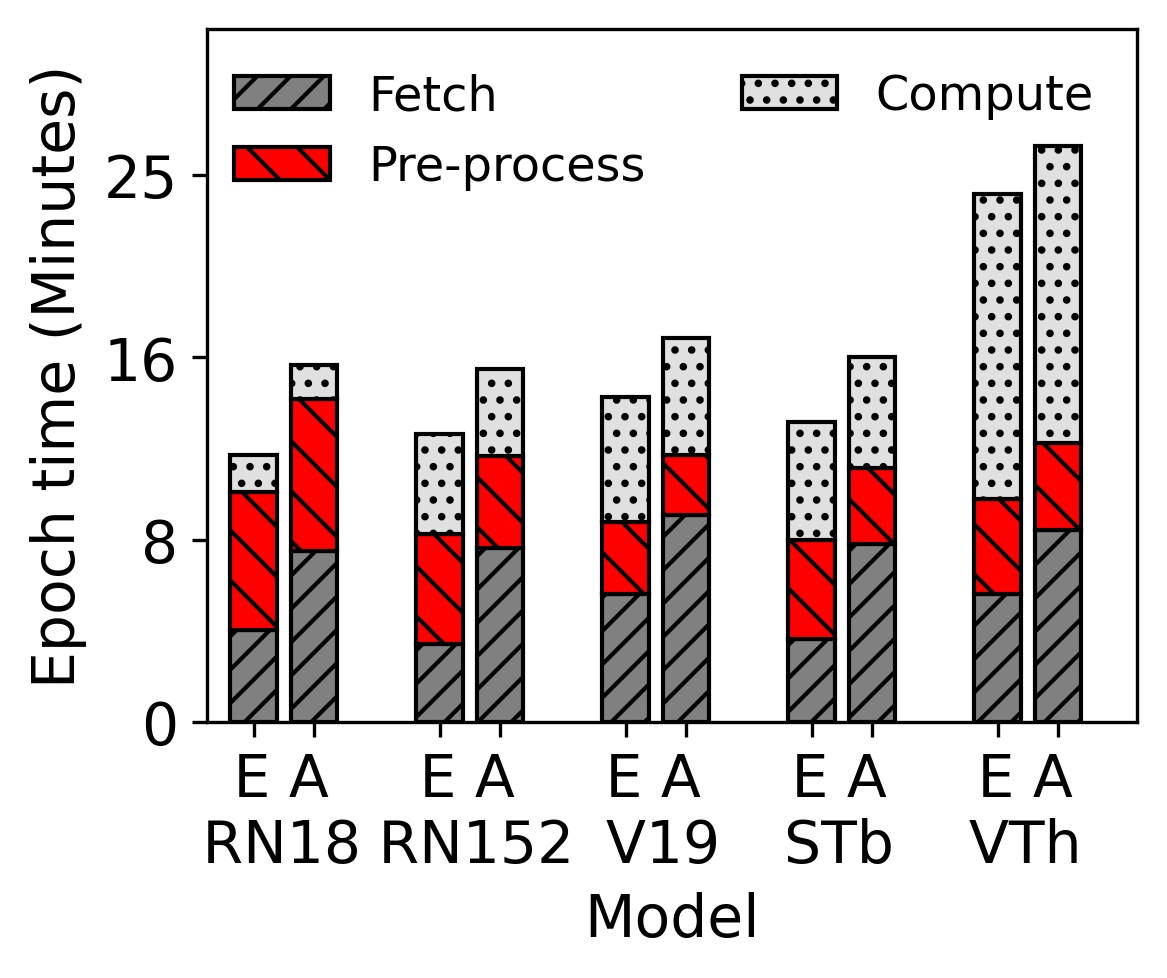}
                \caption{Epoch completion times with 250GB cache}
                \label{fig:fetch_prep_compute_250g}
        \end{subfigure}%
        \caption{
        Fetch, preprocess, and compute times when data is cached in encoded (`E') or augmented (`A') form for ResNet-18 (RN18), ResNet-152 (RN152), VGG-19 (V19), SwinT big (STb), and ViT huge (VTh) with different cache sizes. 
        }
    \label{fig:fetch_prep_compute}
\end{figure}

\begin{table}[b]
    \centering
    \small
    \caption{Data forms and their trade-offs.
    }
    \begin{tabular*}{\columnwidth}{@{\extracolsep{\fill}}lccc@{}}
        \toprule
        
        & \makecell[cc]{Data \\ density}
        & \makecell[cc]{Training \\ readiness}
        & \makecell[cc]{Cache \\ worthiness} \\
        \midrule
        Encoded & \textcolor{ForestGreen}{High} & \textcolor{red}{Low} &  \textcolor{ForestGreen}{High}\\
        Decoded & \textcolor{red}{Low} & \textcolor{orange}{Medium} & \textcolor{ForestGreen}{High} \\
        Augmented & \textcolor{red}{Low} & \textcolor{ForestGreen}{High} & \textcolor{red}{Low} \\
        \bottomrule
    \end{tabular*}
    \label{tab:caching_forms}
\end{table}

Training data can be cached in different forms (encoded, decoded, or augmented), but choosing the right form involves a space-time trade-off, making it challenging to determine the optimal choice.
Table~\ref{tab:caching_forms} illustrates the trade-off in caching encoded, decoded, and augmented data across three different metrics. 
In terms of data density, the encoded data is the highest (smallest in size), thus for a given capacity, more encoded data can be cached. 
Training readiness represents how preprocessed the data is: for this metric, the final augmented data is the most training-ready. 
Cache worthiness qualifies how useful it is to cache the data. 
While encoded and decoded data can be reused across epochs, repeatedly using the same randomly augmented data risks overfitting due to insufficient randomness in the augmentations.


We illustrate the subtlety of caching in the DSI pipeline by measuring the performance of five models when caching either encoded or augmented data at two different cache capacities (Figure~\ref{fig:fetch_prep_compute}). We add Redis~\cite{redis} to PyTorch~\cite{pytorch} for caching either encoded or augmented data of the OpenImages dataset~\cite{OpenImages} on a CloudLab~\cite{duplyakin2019design} system with 4$\times$A100 GPUs, 2$\times$24 core AMD 7413 CPUs, 512 GB DRAM, 200 Gbps Mellanox ConnectX-6 NIC, and a remote storage service (NFS) for datasets. The measurements show the time spent fetching data, preprocessing on the CPU, and computing on the GPU. Caching augmented data (`A') reduces preprocessing time since only data from storage needs processing, whereas caching encoded data (`E') requires preprocessing all data. However, caching augmented data (`A') increases fetch time due to the larger tensor size, which limits the number of samples stored in cache.



Interestingly, the trade-off between caching strategies depends on cache size. With 450GB (Figure~\ref{fig:fetch_prep_compute_450g}), caching preprocessed data significantly reduces preprocessing time by 69.91\%, while fetch time only increases by 34.85\% on average. However, with 250GB (Figure~\ref{fig:fetch_prep_compute_250g}), the benefit of caching preprocessed data is minimal: preprocessing time decreases by just 11.36\%, while fetch time rises by 87.2\% on average.


While Figure~\ref{fig:fetch_prep_compute} shows the performance impact of caching encoded vs. augmented data based on cache size, other hardware factors also play a role. A faster CPU reduces preprocessing time, and faster storage shortens data fetch time. While we only compare caching encoded or augmented data, a range of options exists, such as caching decoded data or splitting the cache between different forms. Deciding how to best allocate a limited cache for the DSI pipeline is a complex, non-linear problem influenced by multiple variables.

\subsection{Caching for random sample accesses}

\begin{figure}[t!]
    \begin{subfigure}[t!]{0.46\columnwidth}
        \includegraphics[width=\linewidth]{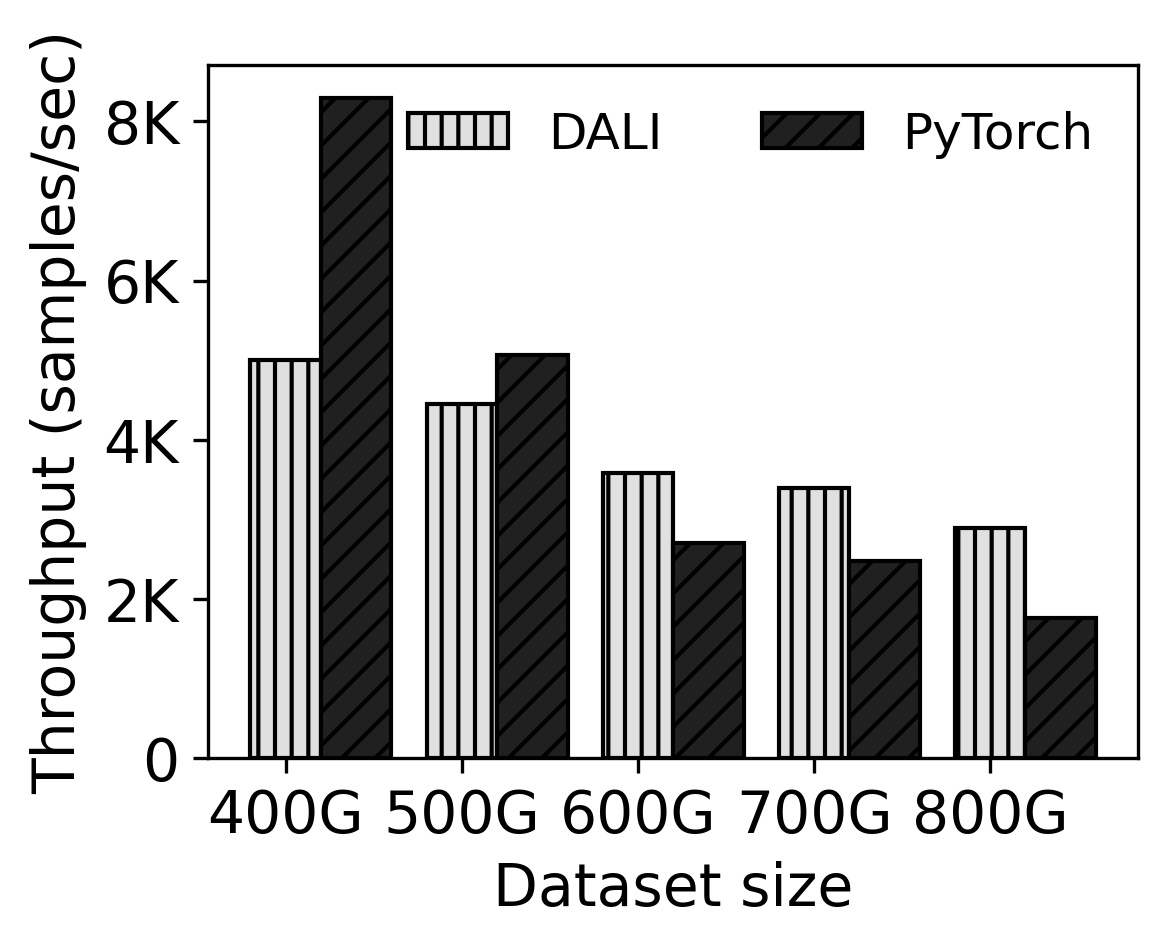}
        \caption{
        DSI throughput with respect to varying dataset size (in GB) for popular open-source dataloaders. 
        }
        \label{fig:dataloader_throughput_memory}
    \end{subfigure}%
    \hfill
    \begin{subfigure}[t!]{0.49\columnwidth}
        \includegraphics[width=\linewidth]{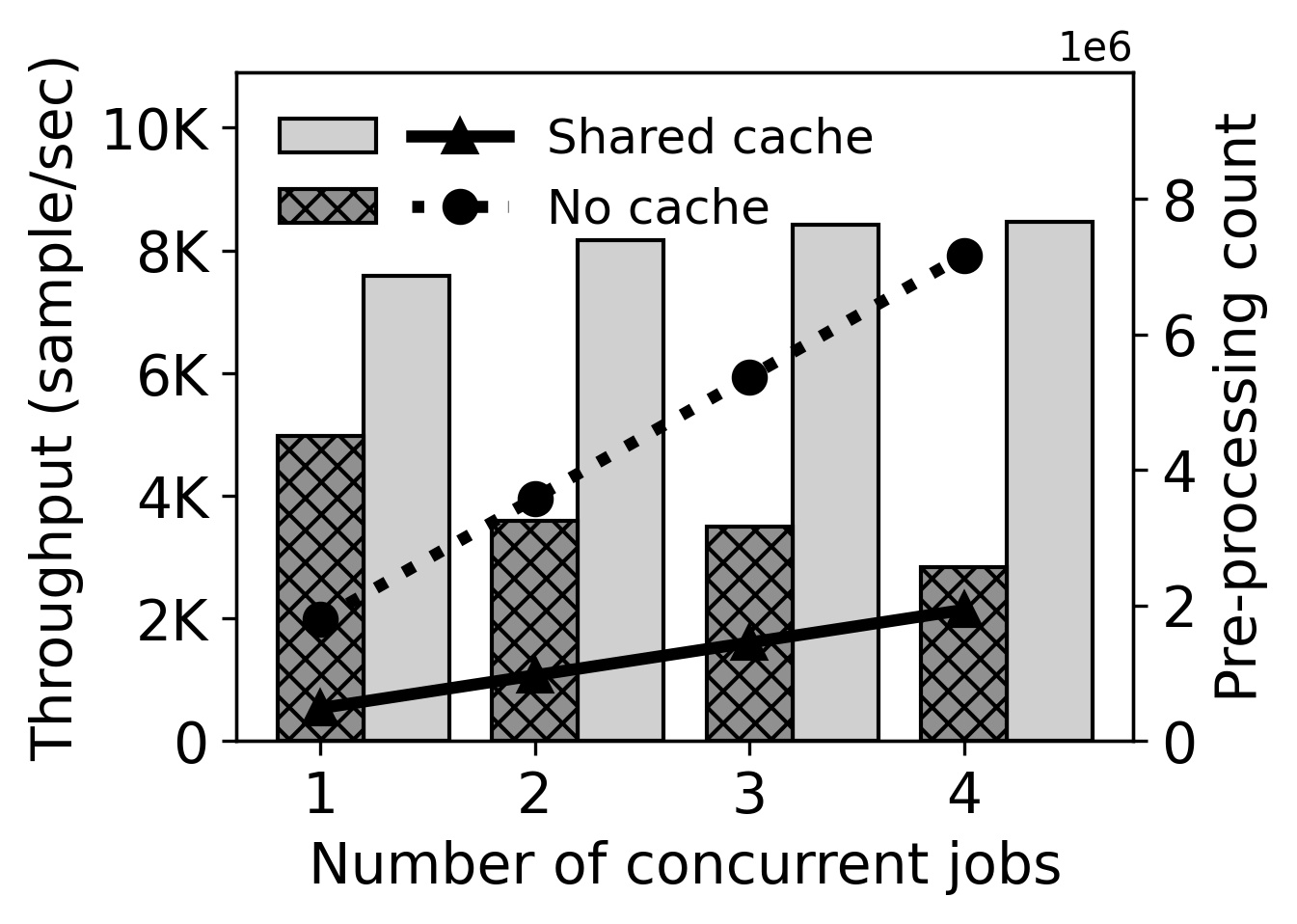}
        \caption{
        DSI throughput wrt. number of jobs with and without caching. Lines: preprocessing count; Bars: DSI throughput.
        }
        \label{fig:dataloader_throughput_multijob}
    \end{subfigure}
    \caption{
    Drawback of OS page-cache for random access patterns (Figure~\ref{fig:dataloader_throughput_memory}) and the impact of sharing preprocessed data with concurrently training jobs (Figure~\ref{fig:dataloader_throughput_multijob}). 
    }
    
    \label{fig:br_motivation}
\end{figure}

The inherent nature of random sampling makes it difficult to implement effective caching policies 
and we observe that multiple training jobs that share the same dataset do not synergize. 
In Figure~\ref{fig:br_motivation}, we demonstrate the effect of DRAM size and number of concurrent jobs 
by measuring the DSI throughput when training ResNet-50~\cite{he2016deep} for image classification on the CloudLab system described in \S\ref{sec:motivation_1}.

Figure~\ref{fig:dataloader_throughput_memory} compares two open-source dataloaders, PyTorch~\cite{pytorch} and DALI~\cite{nvidia-dali}, both reliant on the system-wide page cache. The results show that the page cache’s LRU-based algorithm performs poorly with random access patterns. As the dataset size grows, more data is fetched from slow remote storage, degrading DSI pipeline performance. Increasing the dataset size from 400GB to 600GB reduces DSI throughput by 28.41\% for DALI and 67.34\% for PyTorch. While PyTorch outperforms DALI when the dataset fits in page cache, DALI’s more efficient cache usage gives it an advantage as dataset size increases.


Figure~\ref{fig:dataloader_throughput_multijob} illustrates inefficiencies in concurrent training, where each job redundantly preprocesses images independently. Training four concurrent PyTorch jobs without caching leads to 7.16 million total preprocessing operations for 1.7 million OpenImages~\cite{OpenImages} samples (dotted line). As a result, increasing jobs from one to four reduces total DSI throughput by 46.80\% (black bars), while GPU utilization remains below 80\%.  
To demonstrate the benefits of shared caching, we add a 350GB Redis cache with PyTorch to store and share preprocessed data. With a portion of the dataset in cache, preprocessing operations drop 3.7$\times$ (solid line), and aggregate training throughput improves by 11.81\% (gray bars).  
While sharing preprocessed data reduces redundant preprocessing, performance gains are marginal and jobs fail to benefit from higher concurrency. A more effective sampling policy that optimizes cache utilization could further enhance performance.

\section{Design and Implementation of Seneca}

We now describe the two key components of our work: (1) Model-Driven Partitioning (MDP)---a cache partitioning scheme that uses a high-level performance model of the DSI pipeline to determine the optimal ratio of cache to be allocated to each form,
and (2) Opportunistic Data Sampling (ODS)---a data sampler that prioritizes serving cached data over uncached ones to improve cache hit rate and multi-job DSI throughput.

\subsection{Model-driven partitioning (MDP)} \label{sec:mdp}

\begin{figure}
    \centering
    \includegraphics[width=\columnwidth]{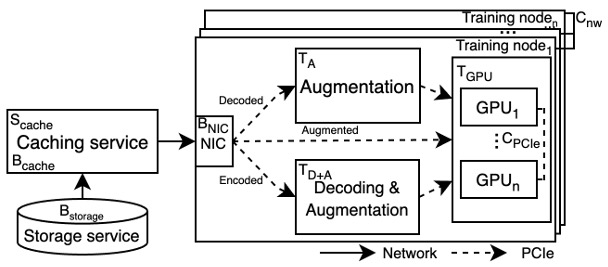}
    \caption{
        The DSI pipeline model.
    }
    \label{fig:dsi_model_cases}
\end{figure}

\begin{table}[!b]
\small
\centering
\caption{Summary of parameters used in the model.
}
{ 
\renewcommand{\arraystretch}{1.2}
\begin{tabular}{|l|l|}
\hline
{\bf Notation} & {\bf Description}\\ \hline
$\mathrm{T}_{GPU}^{}$ & \makecell[ll]{Per node GPU ingestion throughputs \\(sample/s)}  \\ \hline
$\mathrm{T}_{D+A}^{}$, $\mathrm{T}_{A}^{}$ & \makecell[ll]{Per node CPU throughputs for decoding \\and augmenting data (sample/s)} \\ \hline
$\mathrm{B}_{PCIe}^{}$ & \makecell[ll]{Per node PCIe bandwidth \\(B/s)} \\ \hline
$\mathrm{B}_{cache}$ & \makecell[ll]{Maximum remote cache bandwidth\\ (B/s)} \\ \hline
$\mathrm{B}_{storage}$ & \makecell[ll]{Maximum remote storage bandwidth\\ (B/s)} \\ \hline
$\mathrm{B}_{NIC}$ & \makecell[ll]{Per node network bandwidth \\(B/s)} \\ \hline
$\mathrm{S}_{cache}^{}$ & \makecell[ll]{Size of remote cache (Bytes)}\\ \hline
$\mathrm{S}_{data}^{}$ & \makecell[ll]{Size of a encoded data sample (Bytes)} \\ \hline
$\mathrm{N}_{E}$, $\mathrm{N}_{D}$, $\mathrm{N}_{A}$ & \makecell[ll]{Number of cached samples in encoded, \\ decoded, and augmented forms} \\ \hline
$\mathrm{N}_{storage}$ & \makecell[ll]{Number of samples in storage} \\ \hline
$\mathrm{N}_{total}$ & \makecell[ll]{Number of samples in the dataset} \\ \hline
$\mathrm{M}$ & \makecell[ll]{Size inflation factor for\\ preprocessed data} \\ \hline
$\mathrm{C}_{PCIe}$ & \makecell[ll]{Intra-node gradient communication\\ overhead (Bytes)} \\ \hline
$\mathrm{C}_{nw}$ & \makecell[ll]{Inter-node gradient communication \\overhead (Bytes)} \\ \hline
$\mathrm{DSI}_{E}^{}$, $\mathrm{DSI}_{D}^{}$, $\mathrm{DSI}_{A}^{}$ & \makecell[ll]{DSI performance for accessing encoded,\\ decoded, and augmented caches} \\ \hline
$\mathrm{DSI}_{S}^{}$ & \makecell[ll]{DSI performance for fetching data \\from storage} \\ \hline
$\mathrm{DSI}_{overall}^{}$ & \makecell[ll]{Overall DSI performance} \\ \hline
$\textit{x}_{E}^{}$,$\textit{x}_{D}^{}$,$\textit{x}_{A}^{}$ & \makecell[ll]{Proportions of memory allocated \\to the three data forms} \\ \hline
$\textit{n}$ & \makecell[ll]{Number of training nodes} \\ \hline

\end{tabular}
}
\label{tab:model_vars}
\end{table}

We present a high-level performance model that estimates the overall performance of the DSI pipeline for an ML training job. Our formulation is based on data parallel training in a typical setup with GPU training nodes along with remote caching and storage services as shown in Figure~\ref{fig:dsi_model_cases}. Each GPU processes a different batch in parallel and model parameters are synchronized after every batch. Our mathematical model estimates DSI throughput given
(1) parameters for the hardware components of the training nodes (such as an abstract CPU, GPU, memory, and network performance), 
(2) performance of the remote cache and storage services (such as cache and storage bandwidth)
(3) parameters for the training job (such as the dataset size, ML model size, and the average sample size), 
(4) gradient communication overhead,
and (5) the proportions of memory allocated to the three data forms (i.e., encoded, decoded, and augmented). 


Table~\ref{tab:model_vars} summarizes the model parameters and their descriptions. The training node parameters ($\mathrm{T}_{GPU}^{}$, $\mathrm{T}_{A}^{}$, $\mathrm{T}_{D+A}^{}$, $\mathrm{B}_{NIC}^{}$, $\mathrm{B}_{PCIe}^{}$) reflect the performance of hardware components in a single node. 
To obtain the maximum performance of a homogeneous \textit{n}-node training cluster, we multiply these values by the number of nodes ($\textit{n}$) in the cluster. Although the performance of cache and storage service ($\mathrm{B}_{cache}$, $\mathrm{B}_{storage}$) may be internally constrained by various components (e.g., storage device, DRAM bandwidth), we abstract them to the maximum achievable bandwidth from a training node. 


Our model accounts for gradient communication overhead that occurs over PCIe and network after every batch to synchronize gradients across all GPUs as it may overlap with preprocessing tasks. 
We calculate the PCIe and network overheads separately using the number of nodes and GPUs per node.
The network and PCIe overhead, \(\mathrm{C}_{nw}\) and \(\mathrm{C}_{PCIe}\) for a batch is given by \(\frac{2 \times (\textit{n} - 1)}{\textit{n}} \times \beta\mathrm{N}\)~\cite{tang2020communication}, where \(\textit{n}\) is the number of GPUs per node in the case of $\mathrm{C}_{nw}$ and the number of nodes in case of $\mathrm{C}_{PCIe}$ and \(\beta\mathrm{N}\) is the model size in megabytes. For NVLink-connected GPUs~\cite{nvlink}, gradient communication incurs no overhead since NVLink, a dedicated interconnect, is used when available. Specifically, for intra-node NVLink, \(\mathrm{C}_{PCIe} = 0\), and for inter-node NVLink, both \(\mathrm{C}_{PCIe}\) and \(\mathrm{C}_{nw}\) are set to 0.
While we estimate the overhead of ring-reduce which is used in this case, $\mathrm{C}_{nw}$ and $\mathrm{C}_{PCIe}$ can represent overheads for other gradient communication algorithms as well.



Given these parameters, we approach the problem of modeling the DSI pipeline by considering four different cases of data accesses:
(1) when the data needed by the DNN training job is already augmented and in cache \S~\ref{sec:mdp_aug};
(2) when the data is decoded and in cache \S~\ref{sec:mdp_dec};
(3) when the data is encoded and in cache \S~\ref{sec:mdp_raw};
and (4) when the data is in storage \S~\ref{sec:mdp_storage}. 
Our high-level model formulates the performance for each case independently, 
and unifies them into one model by considering the probabilities of all cases  \S~\ref{sec:mdp_overall}. 

\subsubsection{Augmented data in memory} \label{sec:mdp_aug}

We first model the performance of accessing augmented data stored in memory ($DSI_A$). 
For ease of calculation, we use throughput (samples per second) for all the units for the system parameters. 
Because GPU performance is nominally expressed in floating point operations per second (FLOPS), we measure the data ingestion rate for the GPU in terms of samples per second.
The components involved in this scenario are cache, GPUs, and the interconnects (NIC and PCIe) between them. 
The limiting factor for $DSI_A$ in an \textit{n}-node training cluster could be the cache bandwidth ($B_{cache}$) or the aggregate
performance of one of the hardware components in the cluster, which includes GPU ingestion rate ($\textit{n}\times T_{GPU}$), network bandwidth ($\textit{n}\times B_{NIC}$), and PCIe bandwidth ($\textit{n}\times B_{PCIe}$), as shown in Equation~\ref{eq:1}.
PCIe and network bandwidths scale down with both, augmented data size and respective gradient communication overheads ($\mathrm{C}_{nw}$ and $\mathrm{C}_{PCIe}$), while cache bandwidth scales down by the size of augmented data ($M \times S_{data}$).


\begin{dmath}\label{eq:1}
    \mathrm{DSI}_{A}^{} = \min\left(
    \frac{\mathrm{B}_{cache}}{\mathrm{M\times S_{data}}} , 
    \frac{\textit{n}\times\mathrm{B}_{NIC}}{\mathrm{(M \times S_{data})+\mathrm{C_{nw}}}} , 
    \right. \\
    \left.
    \frac{\textit{n}\times\mathrm{B}_{PCIe}}{\mathrm{(M \times S_{data})+\mathrm{C_{PCIe}}}} , 
    \textit{n}\times\mathrm{T}_{GPU}^{}\right)
\end{dmath}

To compute the probability of accessing augmented data, we model the number of samples that are in the augmented form ($N_{A}$). 
Given $x_A$ as the portion of memory allocated for caching augmented data (where $0 \le x_A \le 1$), 
the number of augmented samples in memory is ${x}_{A} \times S_{mem}$ divided by 
the size of a tensor, $M \times S_{data}$.
However, $N_{A}$ should also account for the case when the dataset is small enough to fit entirely in memory, as shown in Equation~\ref{eq:5}. 

\begin{dmath}\label{eq:5}
    \mathrm{N}_{A}^{} = \min\left(\mathrm{N}_{total}^{}, \frac{\mathrm{{\it x}_{A}}\times\mathrm{S}_{mem}^{}}{M\times\mathrm{S}_{data}}\right)
\end{dmath}

\subsubsection{Decoded data in memory}  \label{sec:mdp_dec}

Next, we model the performance of accessing decoded data from memory ($DSI_D$). This scenario involves training node CPUs for applying random augmentations, along with GPUs, remote cache, and interconnects (Network and PCIe) between them.
Similar to the GPU, we represent CPU throughput in samples per second for random augmentations, even though its nominal metric is FLOPS.
The limiting factor for $DSI_D$ is either the remote cache bandwidth ($B_{cache}$) or the aggregate performance of one of the hardware components in the cluster, which includes CPU augmentation throughput ($\textit{n}\times\mathrm{T}_{A}$), GPU ingestion rate ($\textit{n}\times\mathrm{T}_{GPU}^{}$), and interconnect throughput ($\textit{n}\times B_{NIC}$, $\textit{n}\times B_{PCIe}$), as shown in Equation~\ref{eq:2}.

\begin{dmath}\label{eq:2}
    \mathrm{DSI}_{D}^{} = \min\left(
    \frac{\mathrm{B}_{cache}}{\mathrm{M \times S_{data}}},
    \frac{\textit{n}\times\mathrm{B}_{NIC}}{\mathrm{(M \times S_{data})+\mathrm{C_{nw}}}},
    \textit{n}\times\mathrm{T}_{A}, \frac{\textit{n}\times\mathrm{B}_{PCIe}}{\mathrm{(M \times S_{data})+\mathrm{C_{PCIe}}}} \textit{n}\times\mathrm{T}_{GPU}^{}
    \right)
\end{dmath}

To compute the probability of accessing decoded data from memory, we model the number of samples that are in the decoded form (${N}_{D}$). Given ${x}_{D}$ as the portion of memory allocated for decoded data, the number of decoded samples in memory is ${x}_{D} \times S_{mem}$ divided by the size of a tensor, $M \times S_{data}$. 
However, $N_D$ should also account for the case when the dataset is small enough so that both $N_D$ and $N_A$ fit entirely in memory, as shown in Equation~\ref{eq:6}. 

\begin{dmath}\label{eq:6}
    \mathrm{N}_{D} = \min\left(\mathrm{N}_{total}-\mathrm{N_A}, \frac{\mathrm{{\it x}_{D}}\times\mathrm{S}_{mem}}{M\times\mathrm{S}_{data}}\right)
\end{dmath}

\begin{figure*}
    \centering
    \includegraphics[width=0.9\textwidth]{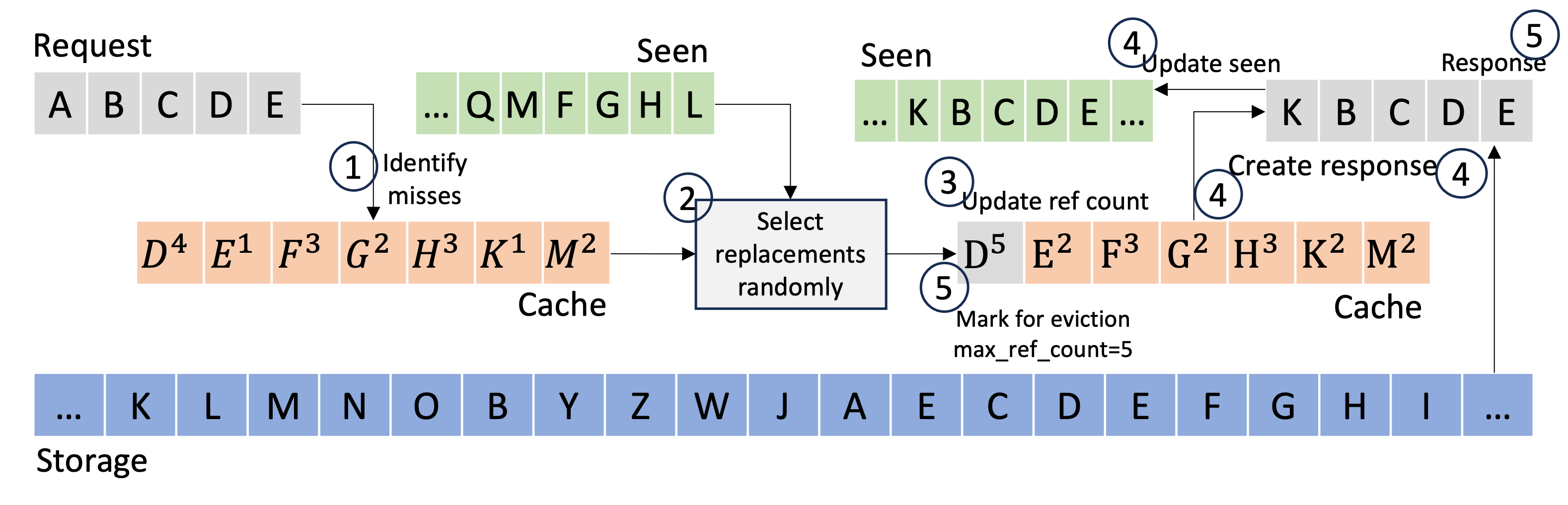}
    \caption{An example operation of ODS. ODS maintains two metadata structures: 
    (1) per-job {\tt seen} bit vector to track whether a data sample has been used during that epoch, 
    and (2) per-dataset {\tt status} and {\tt reference} count for each data sample. 
    These are used to opportunistically replace requested samples that miss in the cache with those that hit while ensuring randomness. 
    }
    \label{fig:cache_aware_sampling}
\end{figure*}

\subsubsection{Encoded data in memory}  \label{sec:mdp_raw}

Now, we model the performance of accessing encoded data stored in memory ($DSI_E$). this involves not only the cache, network, PCIe, and GPU, but also the CPU for decoding and augmenting the data samples ($T_{D+A}$). This resulting $DSI_E$ is shown in Equation~\ref{eq:3}. 

\begin{dmath}\label{eq:3}
    \mathrm{DSI}_{E}^{}=\min\left(\frac{\mathrm{B}_{cache}}{\mathrm S_{data}}, \frac{\textit{n}\times \mathrm{B}_{NIC}}{\mathrm{\mathrm S_{data}+\mathrm{C_{nw}}}}, 
    \textit{n}\times \mathrm{T}_{D+A},
    \frac{\textit{n}\times \mathrm{B}_{PCIe}}{\mathrm{(M \times S_{data})+\mathrm{C_{PCIe}}}}, 
    \textit{n}\times \mathrm{T}_{GPU}^{}\right)
\end{dmath}

For the number of encoded data samples in memory (${N}_{E}$), we consider two scenarios: (1) when the dataset is small enough so that all ${N}_{E}$, $N_D$, and ${N}_A$ can be fully cached, and (2) when the dataset is large. For the latter case, we compute the portion of memory allocated for encoded data, ${x}_{E} \times S_{mem}$, and divide it by the size of the encoded data, $S_{data}$. Both scenarios are considered in Equation~\ref{eq:7}.

\begin{dmath}\label{eq:7}
\mathrm{N}_{E}^{} = \min\left(\mathrm{N}_{total}^{}-(\mathrm{N}_{A}^{}+\mathrm{N}_{D}^{}), \frac{\mathrm{{\it x}_{E}}\times\mathrm{S}_{mem}^{}}{\mathrm{S}_{data}^{}}\right)
\end{dmath}

\subsubsection{Data in storage}  \label{sec:mdp_storage}
Finally, we model the performance of accessing storage, $DSI_{S}$, shown in Equation~\ref{eq:4}.
This is similar to $DSI_E$ in Equation~\ref{eq:3}, but also includes the storage throughput as the potential limiting factor for performance. 

\begin{equation}\label{eq:4}
    \mathrm{DSI}_{S}^{}=\min\left(DSI_E, \frac{B_{storage}}{S_{data}} \right)
\end{equation}

The number of samples only in storage is simply the dataset that does not fit in memory, as shown in Equation~\ref{eq:8}

\begin{equation}\label{eq:8}
    N_{storage} = N_{total} - {N}_{A} -  {N}_{D} - {N}_{E}
\end{equation}

\subsubsection{Overall DSI performance model} \label{sec:mdp_overall}

We combine Equations~\ref{eq:1}--~\ref{eq:8} 
and express the overall DSI throughput ($DSI_{overall}$) as shown in Equation~\ref{eq:9}. 

\begin{equation}\label{eq:9}
\begin{split}
    \mathrm{DSI}_{overall}^{}=\frac{\mathrm{N}_{A}}{\mathrm{\mathrm{N}_{total}^{}}}\times \mathrm{DSI}_{A}^{} + 
    \frac{\mathrm{N}_{D}}{\mathrm{N}_{total}^{}}\times \mathrm{DSI}_{D}^{} + \\
    \frac{\mathrm{N}_{E}}{\mathrm{N}_{total}^{}}\times\mathrm{DSI}_{E}^{}+ \frac{\mathrm{N}_{storage}^{}}{\mathrm{N}_{total}^{}}\times\mathrm{DSI}_{S}^{}
\end{split}
\end{equation}

We use Equation ~\ref{eq:9} to estimate DSI throughput given different proportions of memory partitioned for caching encoded, decoded, and augmented data (${x}_E, {x}_{D}, {x}_A$).
We show the correctness of this model in the later validation section, \S~\ref{sec:valid}.

\subsection{Opportunistic data sampling (ODS)}\label{sec:ods}

Given the optimal cache partitioning for improving the DSI throughput, 
ODS aims to improve the cache hit rate when multiple training jobs share the same dataset. 
The key idea behind ODS is to opportunistically supply data samples already present in the cache. 
However, it must still ensure that 
(1) a training job only sees the same data sample once in any epoch, 
(2) the same augmented samples are never reused across epochs,
and (3) the order of supplied data is random.

We illustrate the operation of ODS through Figure~\ref{fig:cache_aware_sampling}. 
We represent each unique data sample with alphabets, and maintain 
(1) a per-job {\tt seen} bit vector to track whether or not a sample has been seen during its epoch, 
and (2) a per-dataset {\tt status} and {\tt reference} count for each of its data samples. 
Although Seneca has three tiers of caches (augmented, decoded, and encoded), we illustrate with only the augmented cache for simplicity.  

\begin{enumerate}[itemsep=-0.5ex]
    \item When a batch request arrives, ODS identifies the misses (samples not in the cache) based on {\tt status}. 
    \item Using the {\tt seen} bit vector for the requested job, ODS opportunistically replaces misses with hits (samples in the cache) that have not been seen. 
        Hits that have been seen by the job do not replace misses. 
    \item The {\tt reference} counts for samples that hit in the cache are incremented. 
    \item Response to the batch request is sent with replacements, and the {\tt seen} bit vector is updated. 
    \item If any of the {\tt reference} counts reach the defined threshold for eviction, 
        a background thread removes the data samples and replaces them with different random samples from storage with their {\tt reference} counts reset. 
    \item The {\tt seen} bit vector is reset at the end of its epoch.
\end{enumerate}

Using the {\tt seen} bit vector, ODS guarantees that each job uses a data sample once during each epoch. 
With the eviction threshold set to the number of jobs,
the {\tt reference} count and {\tt seen} bit vector together ensure that augmented data will not be used across epochs. 
Although the eviction of an augmented sample from the cache is deterministic, 
when it will be supplied to a job is nevertheless random as it depends on the composition of the requested random samples. 
Furthermore, which data samples populate the augmented cache after an eviction is based on a pseudo-random number generator, warranting the random sampling behavior of ODS. 


The overhead of using ODS is negligible. First, the required metadata for ODS to maintain is only in the megabyte range.
We use 1 bit per data sample for the per-job {\tt seen} bit vector, 
and 1B per data sample for encoding the data status (augmented, decoded, encoded, or storage) and the {\tt reference} count together. 
As an example, training 8 concurrent jobs on the ImageNet-1K dataset~\cite{deng2009imagenet} (1.3M samples) yields 2.6MB of ODS metadata. 
Furthermore, the compute overhead of ODS is limited to only four additional metadata operations, all of which are constant time and in the nanoseconds range. Specifically, these are lookup and update operations on the in-memory ODS metadata.

\subsection{Implementation}

\begin{table}[b]
    \footnotesize
    \centering
    \caption{Hardware configurations for the GPU servers used.}
    \begin{tabular}{l c c c}
        \hline
        & \makecell[cc]{In-house\\server} & \makecell[cc]{AWS\\p3.8xlarge} & \makecell[cc]{Azure\\NC96ads_v4} \\
        \hline
        GPU config & 2$\times$RTX5000 & 4$\times$V100 & 4$\times$A100 \\
        GPU Mem (GB) & 32 & 64 & 320\\
        CPU config & \makecell[cc]{AMD Ryzen \\9 3950X} & \makecell[cc]{Intel Xeon \\E5-2686 v4} & \makecell[cc]{AMD EPYC\\ 7V13} \\
        DRAM cap. (GB) & 115 & 244 & 880 \\
        Network BW. (Gbit/s) & 10 & 10 & 80 \\ 
        NFS BW. (MB/s) & 500 & 256 & 250 \\
        \hline
    \end{tabular}
    \label{tab:hardware_platforms}
\end{table}

\begin{figure}
    \centering
    \includegraphics[width=0.9\columnwidth]{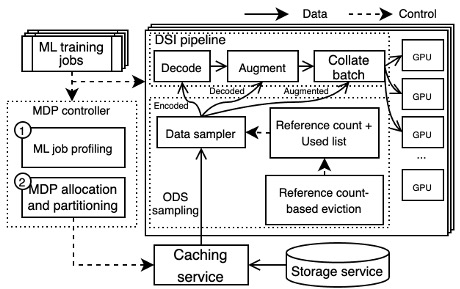}
    \caption{
    The overall architecture of Seneca. 
    At initialization, MDP partitions the cache according to the system and dataset parameters. 
    During runtime, ODS maximizes the cache hit rate by replacing samples that miss in the cache with hits. 
    }
    \label{fig:architecture}
\end{figure}

\begin{figure*}[htbp]
    \centering
    \begin{subfigure}{0.24\textwidth}
        \includegraphics[width=\linewidth]{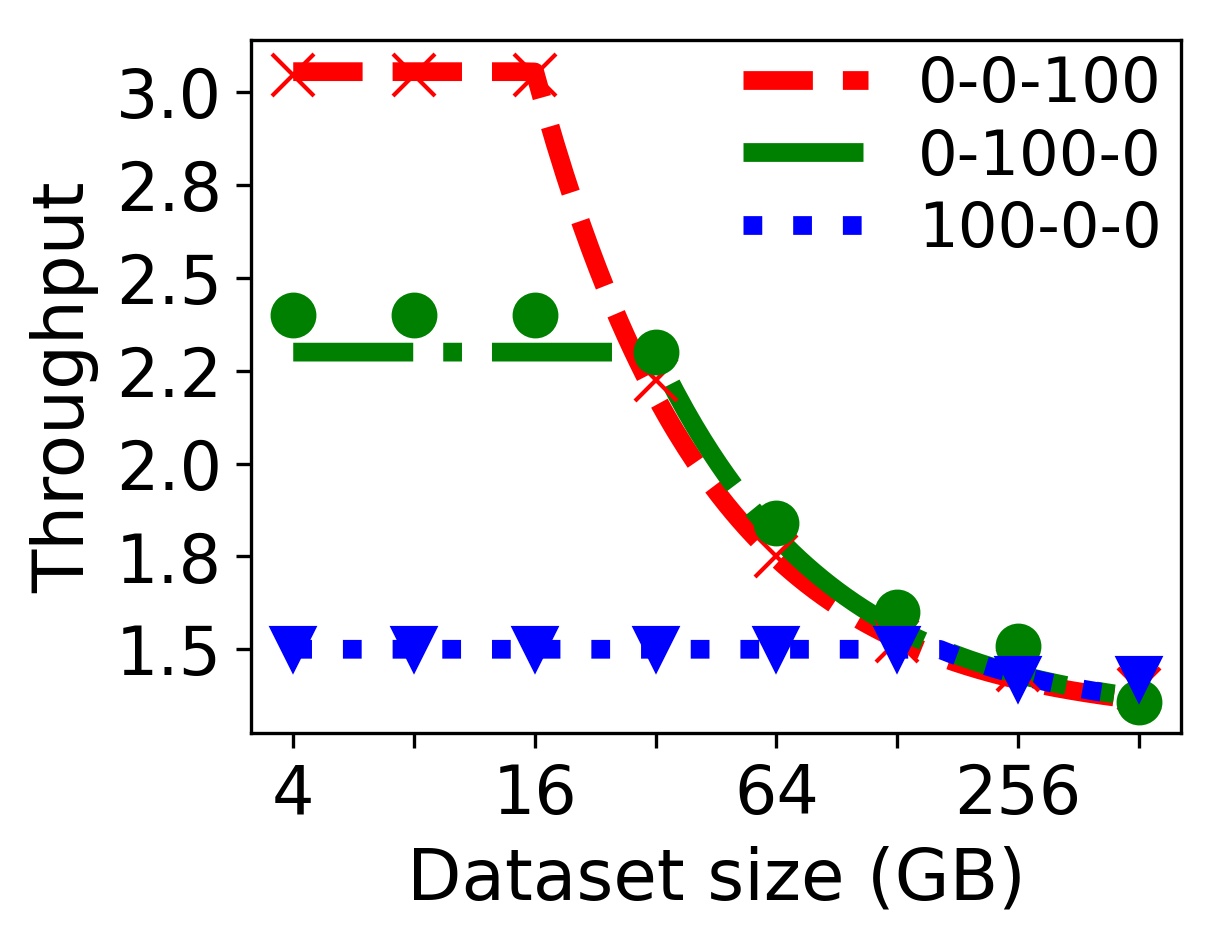}
        \caption{1$\times$in-house, 1$\times$partition}
        \label{fig:hw_a_c1}
    \end{subfigure}
    \begin{subfigure}{0.24\textwidth}
        \includegraphics[width=\linewidth]{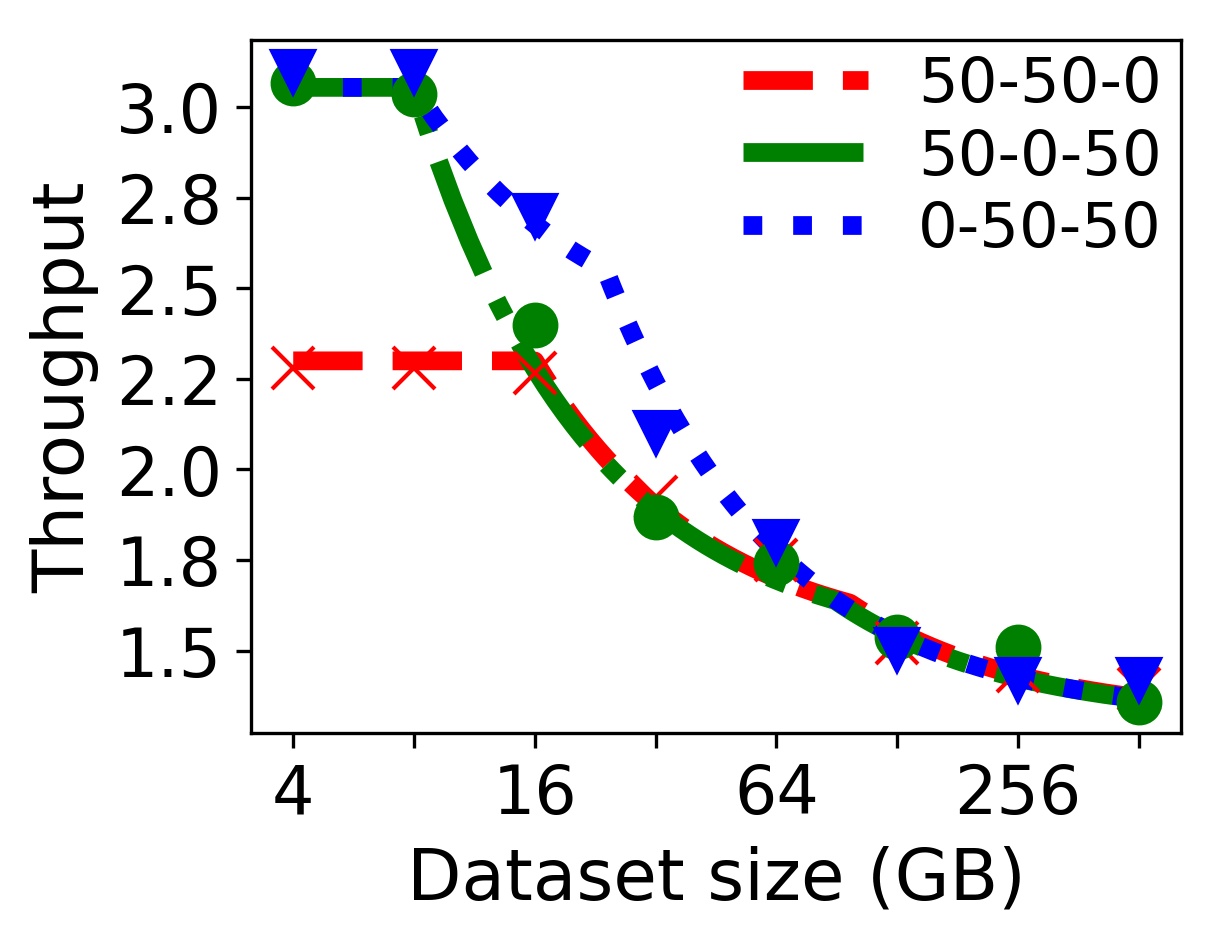}
        \caption{1$\times$in-house, 2$\times$partitions}
        \label{fig:hw_a_c2}
    \end{subfigure}
    \begin{subfigure}{0.24\textwidth}
        \includegraphics[width=\linewidth]{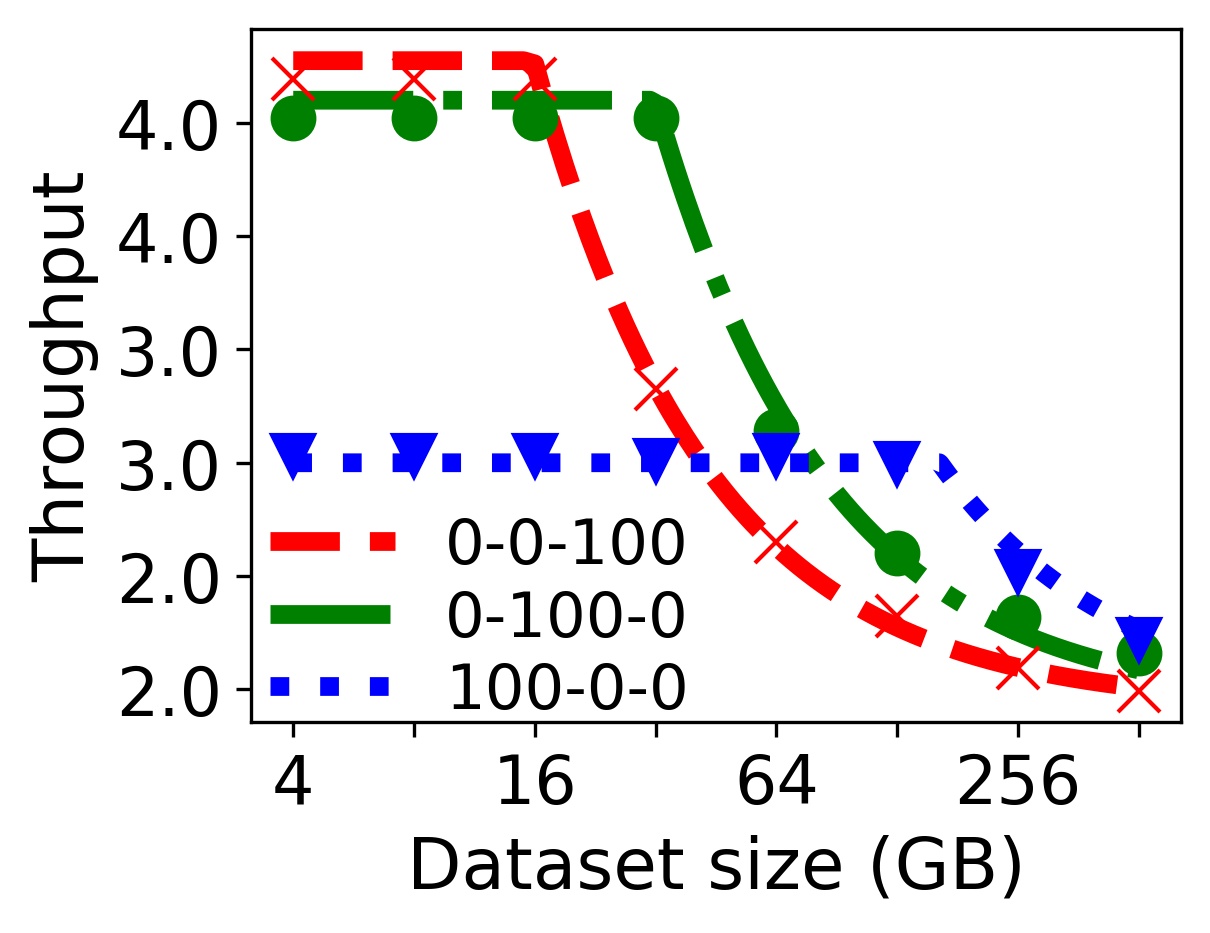}
        \caption{2$\times$in-house, 1$\times$partition}
        \label{fig:hw_a2_c1}
    \end{subfigure}
    \begin{subfigure}{0.24\textwidth}
        \includegraphics[width=\linewidth]{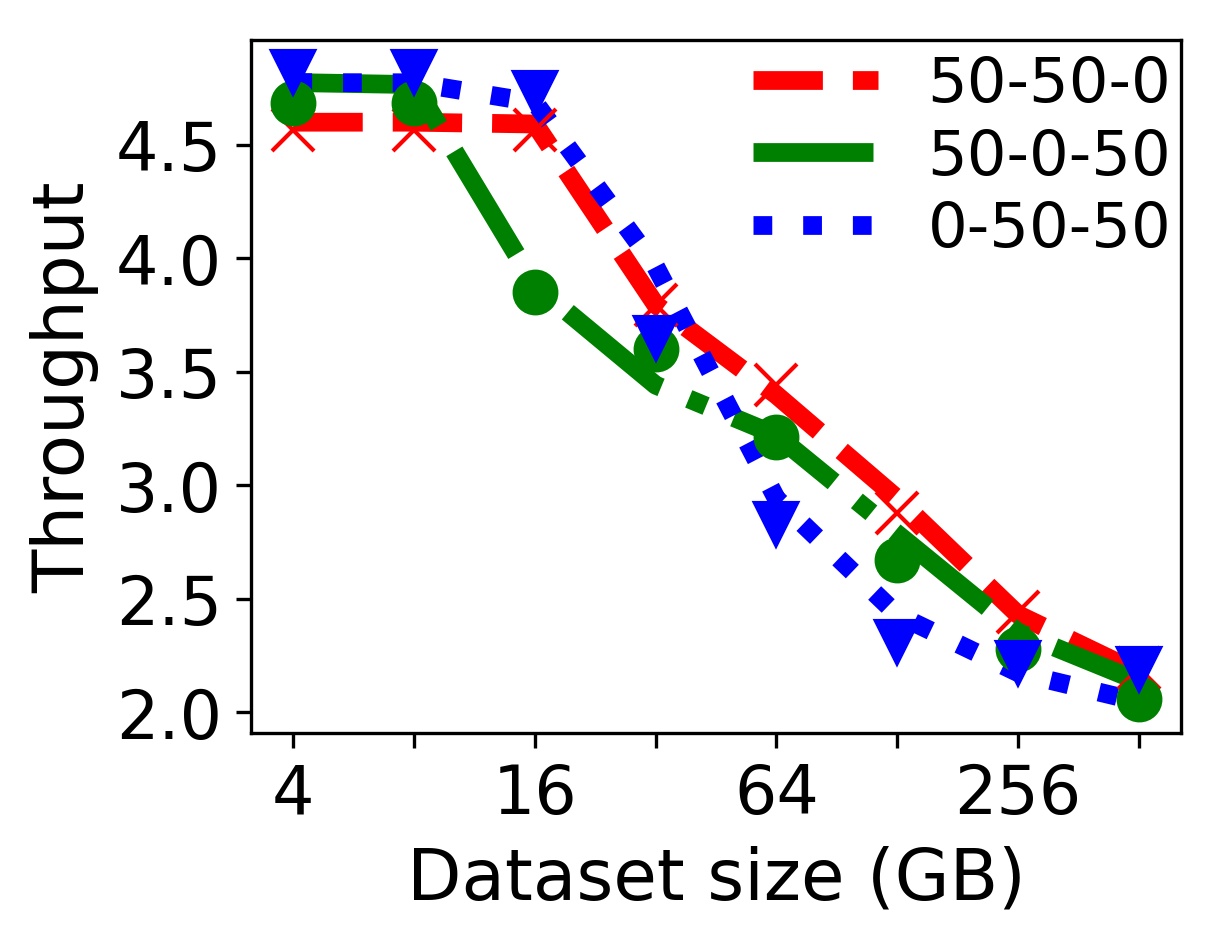}
        \caption{2$\times$in-house, 2$\times$partitions}
        \label{fig:hw_a2_c2}
    \end{subfigure}
    \par\bigskip  
    \begin{subfigure}{0.24\textwidth}
        \includegraphics[width=\linewidth]{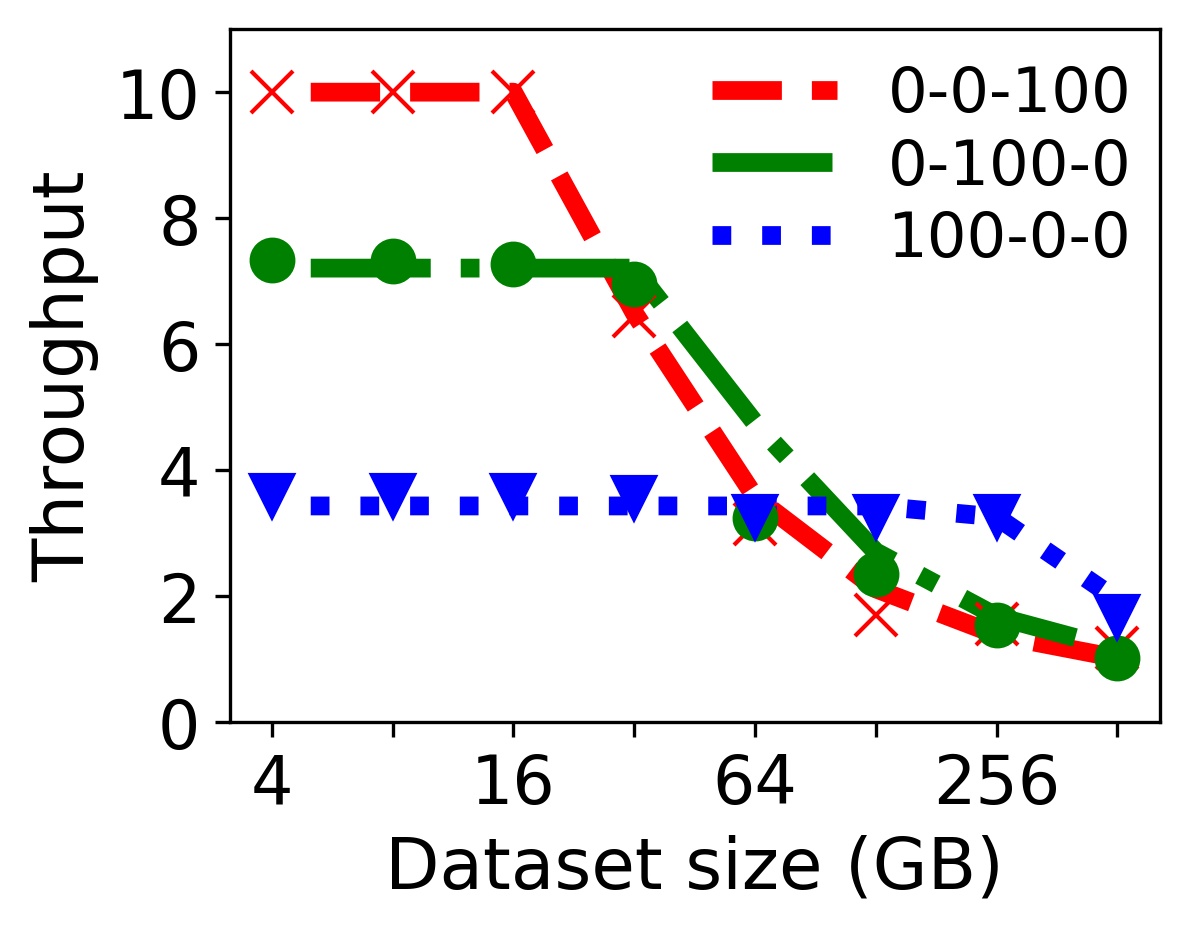}
        \caption{1$\times$AWS, 1$\times$partition}
        \label{fig:hw_p3_c1}
    \end{subfigure}
    \begin{subfigure}{0.24\textwidth}
        \includegraphics[width=\linewidth]{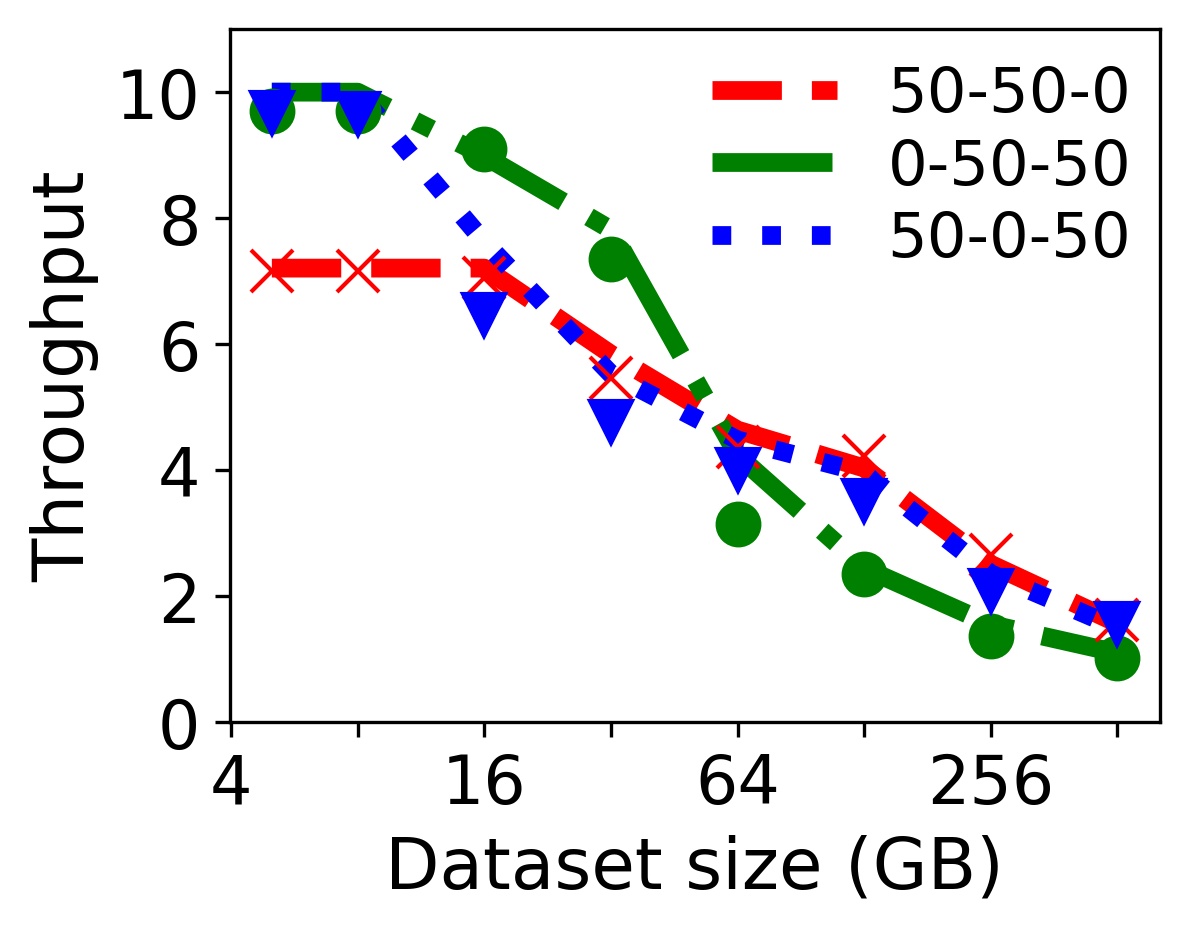}
        \caption{1$\times$AWS, 2$\times$partitions}
        \label{fig:hw_p3_c2}
    \end{subfigure}
    \begin{subfigure}{0.24\textwidth}
        \includegraphics[width=\linewidth]{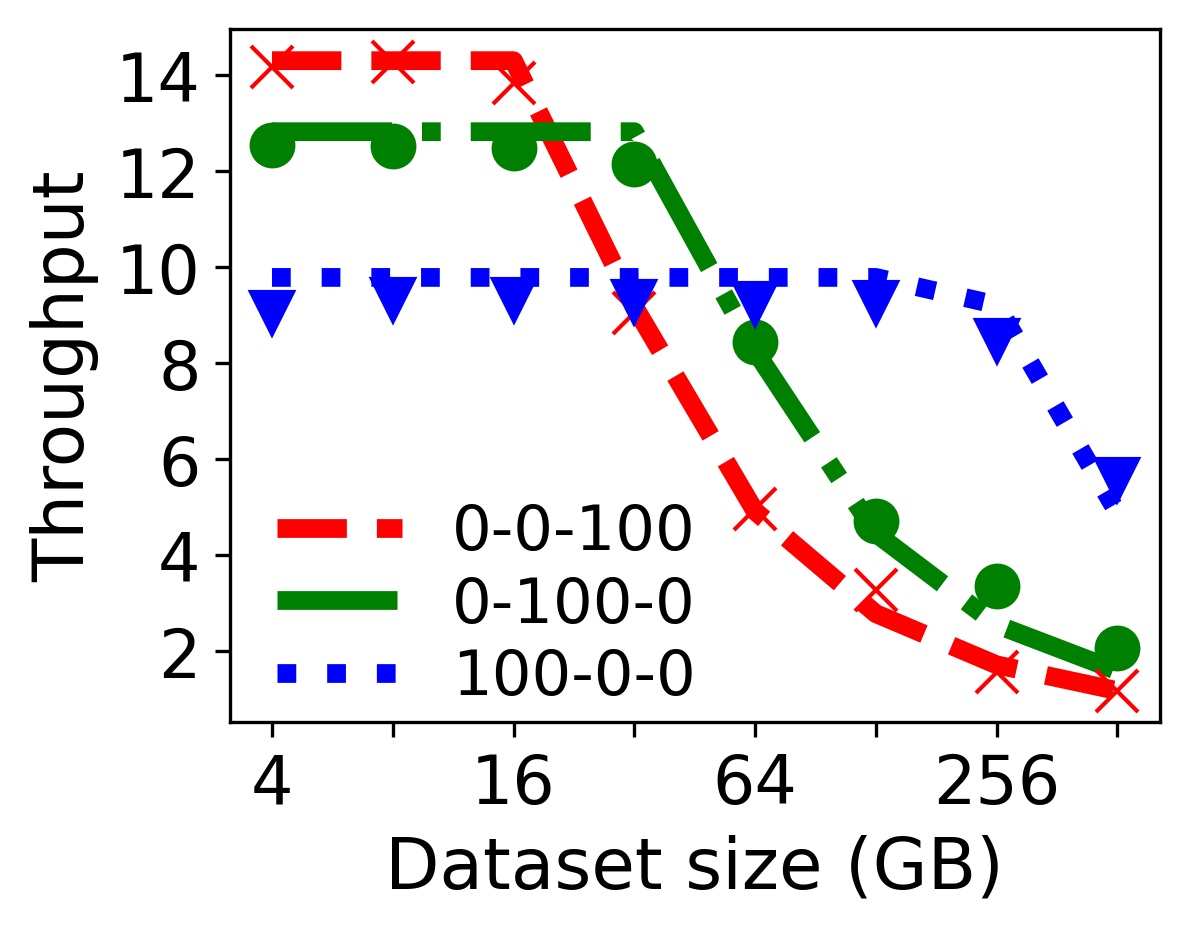}
        \caption{1$\times$Azure, 1$\times$partition}
        \label{fig:hw_azure_c1}
    \end{subfigure}
    \begin{subfigure}{0.24\textwidth}
        \includegraphics[width=\linewidth]{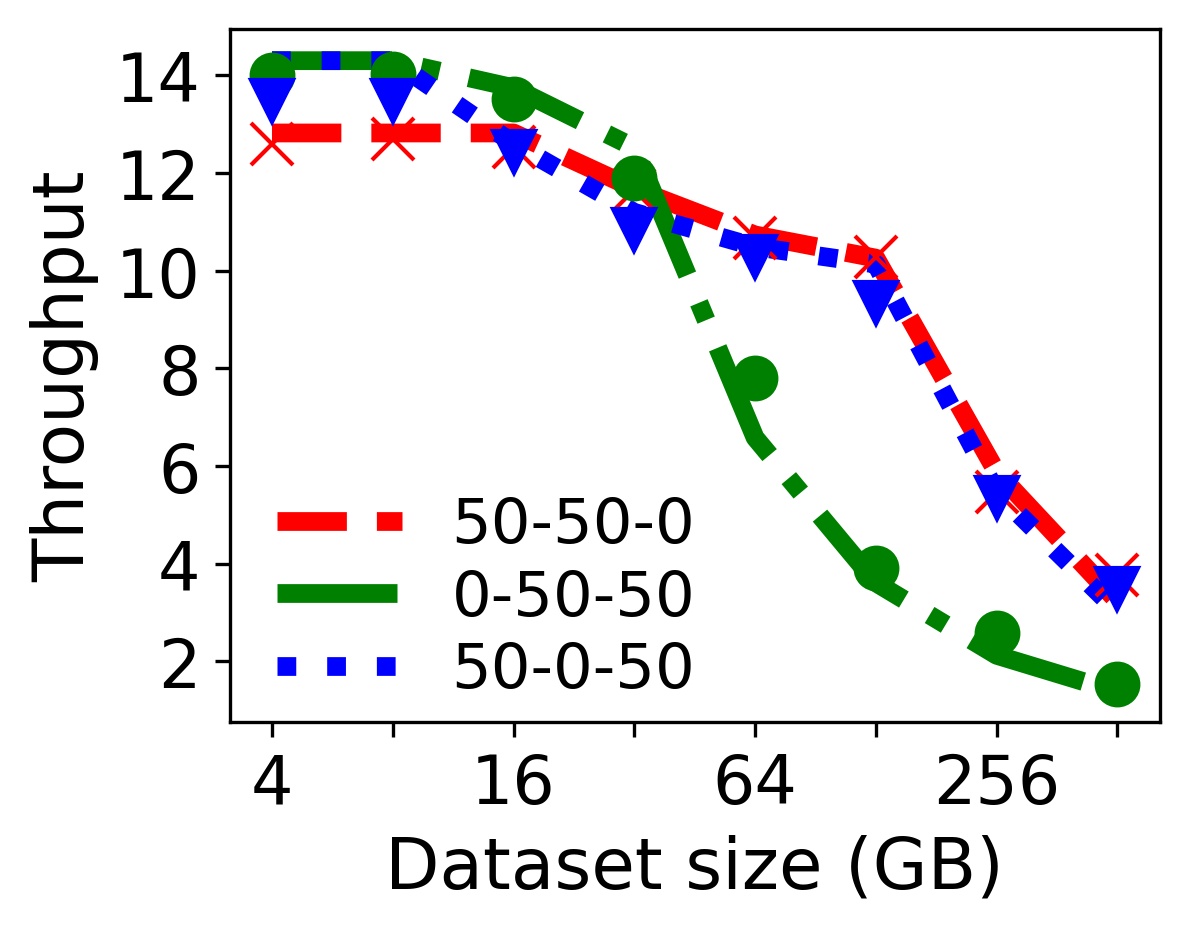}
        \caption{1$\times$Azure, 2$\times$partitions}
        \label{fig:hw_azure_c2}
    \end{subfigure}
    
    \caption{ Validation of the DSI pipeline performance model against measured throughput in samples/sec (Y-axis) while varying the dataset size in GB (X-axis). The lines indicate modeled performance for each cache partition while the markings of the same color indicate measured performance. {\sf X-Y-Z} denote a split of {\sf X}\% encoded cache, {\sf Y}\% decoded cache, and {\sf Z}\% augmented cache. The correlation between modeled and measured values is at least 0.90.}
    \label{fig:model_validation}
\end{figure*}

We implement Seneca by modifying PyTorch (v1.12.0)~\cite{pytorch} dataloaders, with approximately 4200 lines of code changes. 
These changes include both MDP (\S~\ref{sec:mdp}) and ODS (\S~\ref{sec:ods}). 
Seneca can be used as a drop-in replacement for the default PyTorch dataloader. 
For caching, Seneca uses Redis~\cite{redis}, an open-source in-memory key-value store, but other data stores can be used instead.


Figure~\ref{fig:architecture} shows the overall architecture of Seneca. 
At the start of training, MDP determines cache partition sizes and allocates them in the caching service.
We use a brute-force approach to find the optimal cache split by calculating DSI throughput for all combinations at 1\% granularity. 
This approach is used for simplicity since the optimal cache split is typically calculated once per dataset and incurs negligible overhead (<1s). ODS replaces cache misses with hits in runtime and includes modules to track job progress and sample status ensuring randomization and uniqueness of data.

\section{DSI Model Validation}\label{sec:valid}

\begin{table}
\footnotesize
\centering
\caption{Performance model values for DSI model validation.
}
{
\renewcommand{\arraystretch}{1.2}
{
\begin{tabular}{lrrr}
\hline
Notation & In-house server & AWS p3.8xlarge & Azure NC96ads_v4\\ \hline
$\mathrm{T}_{GPU}^{}$ & 4550 sample/s & 9989 sample/s & 14301 samples/s\\ 
$\mathrm{T}_{D+A}^{}$ & 2132 sample/s & 3432 sample/s & 9783 samples/s\\ 
$\mathrm{T}_{A}^{}$ & 4050 sample/s & 6520 sample/s &12930 samples/s\\ 
$\mathrm{B}_{NIC}$    & 10 Gb/s & 10 Gb/s & 80 Gb/s\\
$\mathrm{B}_{PCIe}$    & 32 GB/s & 32 GB/s & 64 GB/s\\
$\mathrm{B}_{cache}$    & 10 Gb/s & 10 Gb/s & 30 Gb/s\\
$\mathrm{B}_{storage}$    & 500 MB/s & 256 MB/s & 250 MB/s\\ 
$\mathrm{S}_{cache}^{}$ & 64 GB & 64 GB & 64 GB\\ 
$\mathrm{S}_{data}^{}$ & 114 KB & 114 KB& 114 KB\\
$\mathrm{M}$ & 5.12$\times$ & 5.12$\times$ & 5.12$\times$ \\ \hline
\end{tabular}
}
\label{tab:nominal_vals}
}
\end{table}


We validate our DSI pipeline performance model by comparing its output with real measurements across four hardware platforms (1$\times$in-house server, 2$\times$in-house servers, 1$\times$AWS server, and 1$\times$Azure server) using several fixed cache partition sizes. The purpose of this validation is to assess the correlation between the model and actual performance, not to demonstrate that MDP achieves optimal results. Table~\ref{tab:hardware_platforms} shows the known hardware specs for the server types used, while Table~\ref{tab:nominal_vals} lists profiled values like GPU and CPU throughput for our model. For each server type, we measure GPU and CPU throughput using DS-Analyzer~\cite{dsanalyzer} and remote storage bandwidth using {\tt fio}~\cite{fio}. We double the performance values of the in-house server for validating our model on distributed training across 2 in-house servers.
Overall, the Azure server generally outperforms the in-house and AWS servers, except in remote storage speed. We use the ImageNet-1K dataset~\cite{deng2009imagenet}, and replicate samples to generate a large dataset that reaches up to 512GB. We also limit the size of the caching service to 64GB on all systems to test scenarios where the dataset exceeds memory capacity.


Figure~\ref{fig:model_validation} compares modeled performance (lines) with measured performance (markings) for six different cache partitions across four configurations. There are countless partition possibilities, so we focus on simple configurations: three with a single cache and three with two equal-sized caches. Modeled and measured results are color-matched and show a strong correlation overall. The Pearson correlation coefficient for all 24 combinations is at least 0.90. The lowest correlations are 0.91 for the 50/50 encoded-augmented cache on the AWS server (blue line in Figure~\ref{fig:hw_p3_c2}) and 0.92 for the 100\% encoded cache on the in-house server (blue line in Figure~\ref{fig:hw_a_c1}). For distributed training on two in-house servers (Figures~\ref{fig:hw_a2_c1} and ~\ref{fig:hw_a2_c2}), the bottleneck shifts to remote cache bandwidth ($\mathrm{B}_{cache}$) due to limited network capacity, a constraint accurately predicted by our model.

\begin{table*}
\centering
\footnotesize
\caption{Characteristics of popular open source datasets used for training image classification models. 
Using the profiled performance model values, MDP determines the best split for the two servers: 
({\sf X-Y-Z}) indicate a split of {\sf X}\% encoded cache, {\sf Y}\% decoded cache, and {\sf Z}\% augmented cache. MDP splits in bold are presented in the evaluation.
}
\begin{tabular}{l c c c c c c c c c c c}
\hline
\multirow{3}{*}{Dataset} & \multirow{3}{*}{Images} & \multirow{3}{*}{Classes} & \multirow{3}{*}{\makecell[cc]{Avg. Image size}} & \multirow{3}{*}{Footprint} & \multicolumn{5}{c}{MDP} \\
\cline{6-10}
 &  &  & & & \makecell[cc]{1$\times$in-house \\server} & \makecell[cc]{2$\times$In-House \\servers} & \makecell[cc]{AWS \\p3.8xlarge} & \makecell[cc]{1$\times$Azure \\NC96ads_v4} & \makecell[cc]{2$\times$Azure \\NC96ads_v4} \\ \hline
ImageNet-1K ~\cite{deng2009imagenet} & 1.3M &1000 & 114.62KB & 142 GB & \textbf{58-42-0} & 40-59-1 & 0-81-19  & \textbf{0-48-52}  & 0-53-47\\
OpenImages V7~\cite{OpenImages} & 1.9M &600 & 315.84KB & 517 GB & \textbf{62-37-1} & \textbf{58-41-1} & \textbf{52-48-0}  & \textbf{5-95-0}  & \textbf{6-93-1}\\  
ImageNet-22K~\cite{imagenet22k} & 14M &22000 & 91.39KB & 1400 GB & 100-0-0 & 100-0-0 & 100-0-0  & \textbf{100-0-0}  & 100-0-0\\ 

\hline
\end{tabular}%
\label{tab:dataset_characteristics}
\end{table*}

\begin{table}
    \centering
    \footnotesize
    \caption{Key features of evaluated dataloaders.
    }
    \begin{tabular*}{\columnwidth}{@{\extracolsep{\fill}}lccc@{}}
        \toprule
        
        & \makecell[cc]{Reduces \\ CPU overhead}
        & \makecell[cc]{Improves \\ cache hit rate}
        & \makecell[cc]{Supports \\ multiple jobs} \\
        \midrule
        PyTorch ~\cite{pytorch} & \xmark & \xmark & \xmark \\
        DALI ~\cite{nvidia-dali} & \cmark & \xmark & \xmark \\
        SHADE ~\cite{khan2023shade} & \xmark & \cmark & \xmark \\
        MINIO ~\cite{mohan2020analyzing} & \cmark & \xmark & \cmark \\
        Quiver ~\cite{kumar2020quiver} & \xmark & \cmark & \cmark \\
        \textit{MDP} & \cmark & \xmark & \cmark \\
        \textit{Seneca} & \cmark & \cmark & \cmark \\
        \bottomrule
    \end{tabular*}
    \label{tab:related}
\end{table}

As shown in the model, the relationship between the DSI pipeline throughput 
and the cache configuration is nontrivial. The DSI pipeline throughput depends on the performance of training node hardware (GPU and CPU), cache and storage services, as well as the dataset and model parameters (size and sample size).
For example, when the dataset is small, then it is advantageous to have preprocessed data in the cache: there is no reason not to as it avoids both preprocessing and I/O stalls as shown by the red lines in Figures~\ref{fig:hw_a_c1}, \ref{fig:hw_a2_c1}, ~\ref{fig:hw_p3_c1}, and ~\ref{fig:hw_azure_c1}. 
However, for larger datasets, the larger size of preprocessed data can lead to performance degradation due to increased cache misses.
If only one type of cache can be used, using an encoded cache is better with large datasets, 
as illustrated by the blue lines and markings in Figures~\ref{fig:hw_a_c1}, \ref{fig:hw_a2_c1}, ~\ref{fig:hw_p3_c1}, and ~\ref{fig:hw_azure_c1}.
With two types of caches, the best configuration is no longer a clear-cut answer, 
although, at very large dataset sizes, their throughputs become similar, shown in Figures~\ref{fig:hw_a_c2}, \ref{fig:hw_a2_c2}, ~\ref{fig:hw_p3_c2}, and ~\ref{fig:hw_azure_c2}.

\section{Evaluation}

We demonstrate the effectiveness of Seneca using several DNN and transformer models with the largest possible batch size up to 1024
on three datasets
across five hardware configurations, incorporating single node and distributed training on a variety of GPUs:
(1) 1$\times$ and 2$\times$in-house servers with 115 GB remote cache over 10 Gbps, (2) AWS p3.8xlarge~\cite{awsp3instancetype} VM with 400 GB remote cache over 10 Gbps on an AWS x2iedn.4xlarge~\cite{awsx2i} VM, and (3) 1$\times$ and 2$\times$Azure NC96ads_v4~\cite{AzureNCv4} VMs with 400 GB remote cache over 30 Gbps on an Azure E64-16s v6~\cite{azureE} VM.
Datasets are stored on a remote storage service, which in our case is an NFS server with 10-12 Gbps network bandwidth.
We use image classification models due to their substantial preprocessing overhead, 
but the core concepts of our work can be applied to all preprocessing-heavy machine learning training. 
The characteristics of the datasets are summarized in Table~\ref{tab:dataset_characteristics} and represent a wide range of dataset sizes and sample sizes. 
How the cache should be partitioned depends on the system and dataset parameters, 
and Table~\ref{tab:dataset_characteristics} includes the MDP-determined partitions (encoded-decoded-augmented). 
Due to space constraints, we present results only for unique MDP partitions. 

We evaluate two of our designs, MDP-only and Seneca (which includes both MDP and ODS), against five dataloaders. We provide a brief description for each of the compared dataloaders and summarize their key features in Table~\ref{tab:related}. 

\begin{description}
    \item[PyTorch] is a popular open-sourced dataloader~\cite{ paszke2019pytorch, pytorch}. 
    \item[DALI] is an open-sourced optimized library that pipelines and offloads preprocessing tasks to the GPU~\cite{nvidia-dali}. 
    \item[SHADE] caches and preferentially samples data with higher importance~\cite{khan2023shade}. 
    We use the unmodified version of SHADE that is publicly available. 
    \item[MINIO] is originally built on top of DALI and does not evict samples once in the cache~\cite{mohan2022looking}. 
    We implement this same policy on top of PyTorch for evaluation. 
    \item[Quiver] samples 10$\times$ more data and forms a batch with those that return the fastest~\cite{kumar2020quiver}. 
    Quiver is not open-sourced, and we faithfully implement Quiver on top of PyTorch. 
   
\end{description}

We answer the following questions in the evaluation. 
\begin{itemize}
\item Do models converge faster with Seneca? (\S~\ref{sec:eval_accuracy})
\item Does Seneca scale with hardware resources?(\S~\ref{sec:hw_scalability})
\item Does Seneca scale with the number of training jobs?(\S~\ref{sec:sw_scalability})
\item How does Seneca perform on different models and datasets? (\S~\ref{sec:epoch_time})
\end{itemize}

\subsection{End-to-end training performance}\label{sec:eval_accuracy}
\begin{figure}
        \begin{subfigure}{0.50\columnwidth}
                \includegraphics[width=\columnwidth]{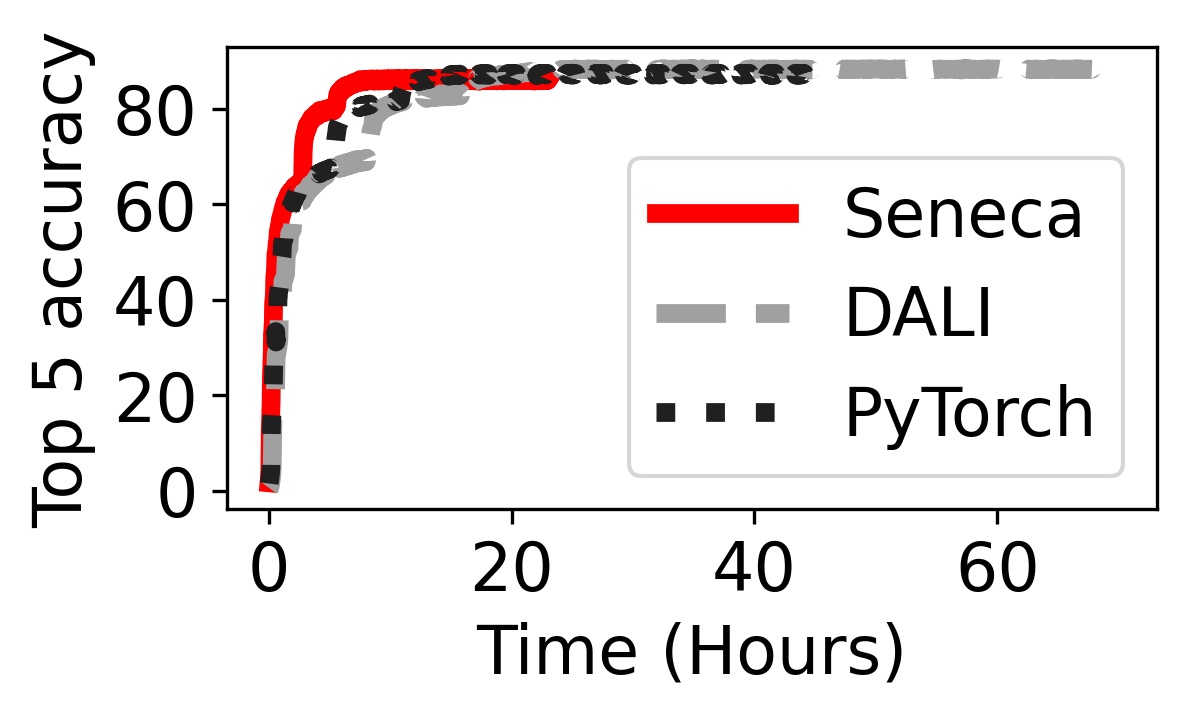}
                \caption{ResNet-18}
                \label{fig:resnet18_accuracy}
        \end{subfigure}%
        \hfill
        \begin{subfigure}{0.50\columnwidth}
                \includegraphics[width=\columnwidth]{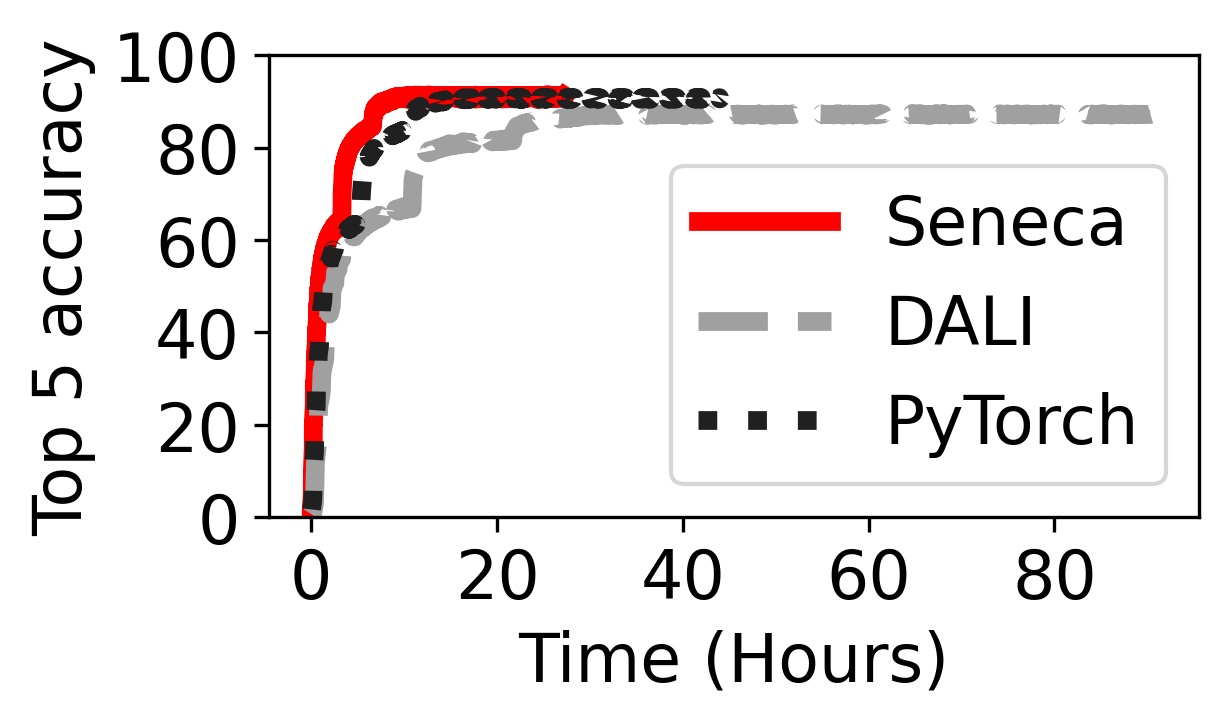}
                \caption{ResNet-50}
                \label{fig:resnet50_accuracy}
        \end{subfigure}
        
        \begin{subfigure}{0.50\columnwidth}
                \includegraphics[width=\columnwidth]{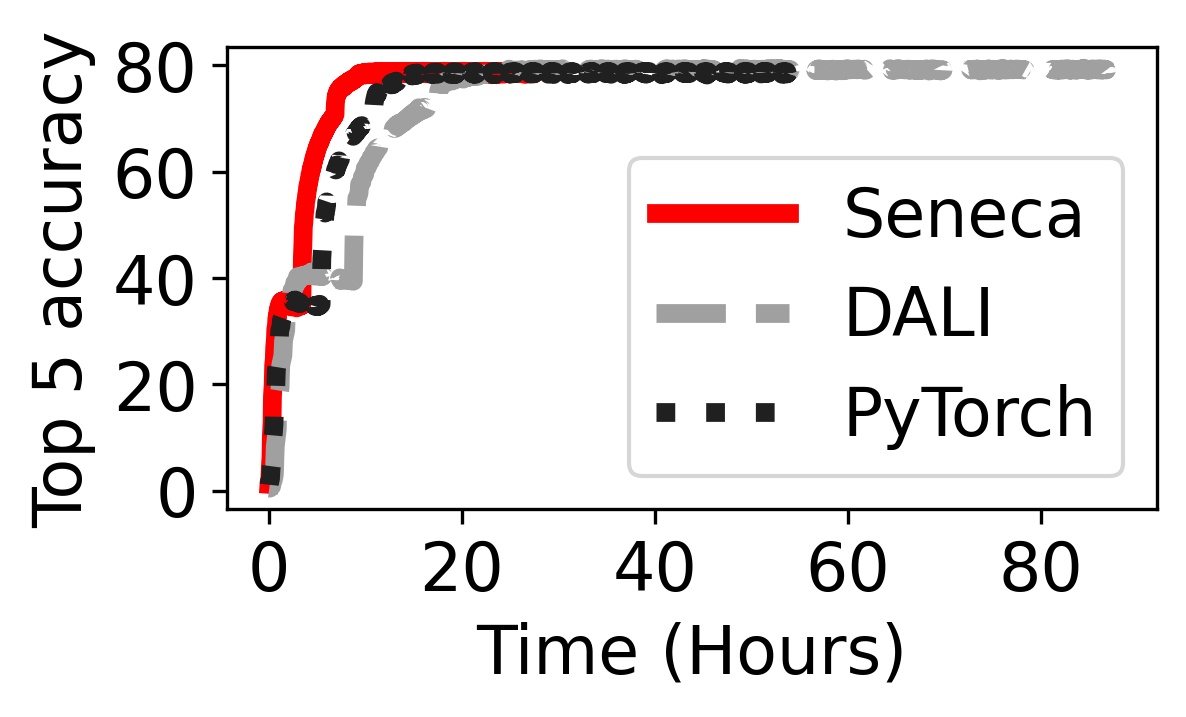}
                \caption{VGG-19}
                \label{fig:vgg19_accuracy}
        \end{subfigure}%
        \hfill
        \begin{subfigure}{0.50\columnwidth}
                \includegraphics[width=\columnwidth]{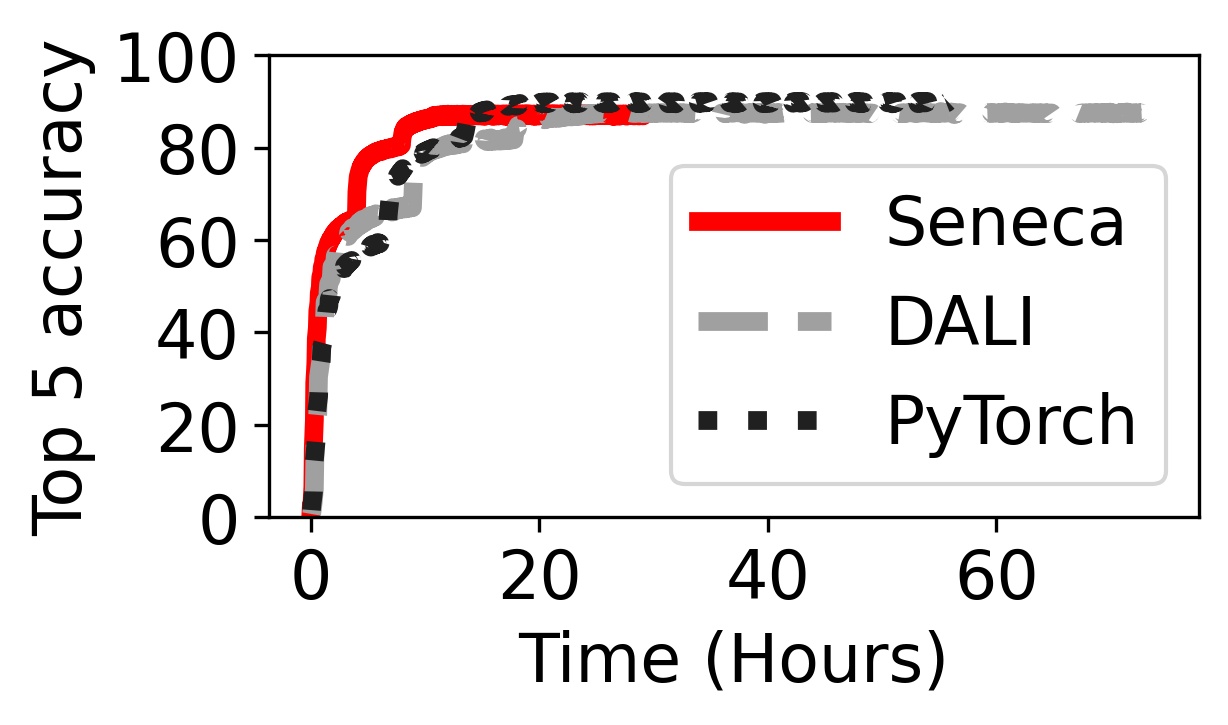}
                \caption{DenseNet-169}
                \label{fig:densenet_accuracy}
        \end{subfigure}
        \caption{Top-5 accuracy of training four ML models on the Imagenet-1K dataset using the Azure VM. 
        Seneca completes 250 epochs significantly faster without compromising accuracy. 
        }\label{fig:accuracy}
\end{figure}

We demonstrate how four popular DNN model architectures converge by 250 epochs using Seneca on the Azure server, as illustrated in Figure~\ref{fig:accuracy}. 
For this evaluation, we choose two GPU-intensive (VGG-19 and DenseNet-169) and two less GPU-intensive models (ResNet-18 and ResNet-50) training on ImageNet-1K which is a dataset commonly used to evaluate image model accuracies~\cite{he2016deep, pinto2018hoard, krizhevsky2012imagenet}.
We choose PyTorch and DALI for comparison as training to convergence takes a considerable amount of time 
and the two open-sourced dataloaders are widely used and stable. 
We show the top-5 accuracy with respect to the time spent training.
First, we observe that Seneca enables DNN models to achieve convergence accuracies significantly faster than PyTorch and DALI. Second, models trained with Seneca follow the same trend in accuracy as those trained with PyTorch and DALI, with an error of less than 2.83\% in the final training accuracy. This demonstrates that our optimizations (MDP and ODS) do not impact on training accuracy.

Next, we investigate time to convergence by measuring 
the end-to-end model training time for the DNN models in Figure~\ref{fig:accuracy}. 
Compared to the PyTorch dataloader, Seneca enables these models to complete the 250 epochs and achieve accuracies of 86.1\% for ResNet-18, 90.82\% for ResNet-50, 78.78\% for VGG-19, and 89.05\% for DenseNet-169, with respective speed improvements of 48.51\%, 38.09\%, 49.16\%, and 47.83\%. When compared against DALI, the improvements are 66.88\%, 70.00\%, 68.53\%, and 60.70\% respectively. For single-job training, Seneca outperforms the other two dataloaders, thanks to MDP which enables caching samples in decoded and augmented formats.



\begin{figure}
\centering
        \includegraphics[width=\columnwidth]{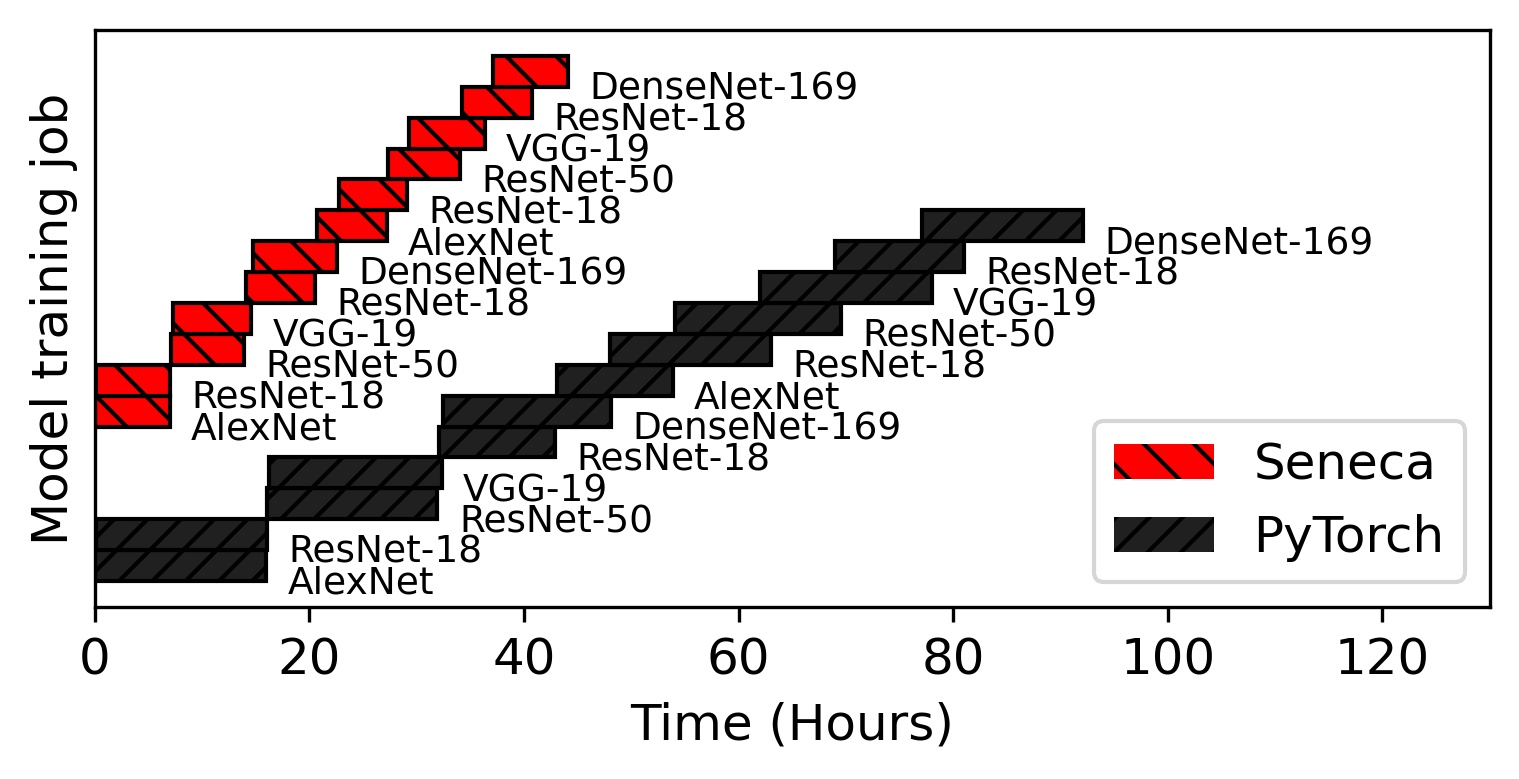}
  \caption{
        The model training time for 12 image classification jobs (50 epochs each) for 5 different models on the AWS server~\cite{awsp3instancetype}.
        For the multi-job training, Seneca reduces the training time by 45.23\% compared to PyTorch.
  }
  \label{fig:makespan}
\end{figure}

We also simulate a real-world training environment on the AWS server using a scheduler to launch jobs arriving at random times. We limit the number of concurrent jobs to two. Figure~\ref{fig:makespan} shows the training progress for 12 jobs (a mix of large and small models) on the ImageNet-1K dataset sharing the same DSI pipeline, each trained for 50 epochs. 
We make three key observations: (1) Seneca reduces the total training time to 45.23\% of PyTorch. (2) Unlike PyTorch, where each job has an independent data pipeline, Seneca's caching system allows data sharing, reducing redundant fetch and preprocessing operations. (3) the last job (DenseNet-169) completed in 6.97 hours compared to the sixth job (DenseNet-169) which completed in 7.84 hours since it ran alone during the second half, fully utilizing the DSI resources.

\vspace{7pt}
\subsection{Hardware sensitivity}\label{sec:hw_scalability}

\begin{figure}
\centering
        \includegraphics[width=\columnwidth]{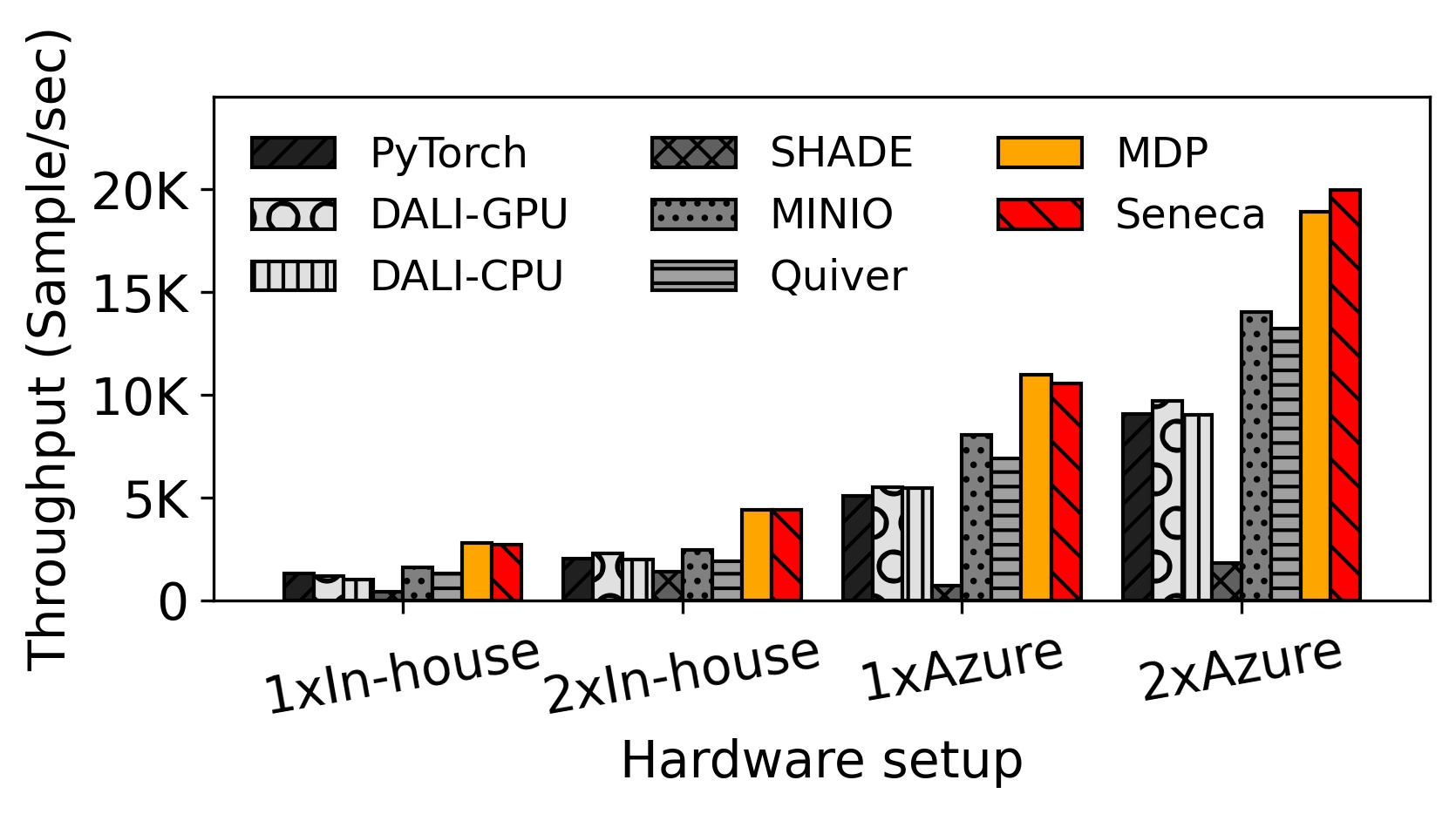}
  \caption{
        Throughput for single job distributed training on one and two in-house and Azure servers. On two Azure nodes, Seneca is 1.89$\times$ faster than one.
  }
  \label{fig:distributed}
\end{figure}

We evaluate Seneca's performance across different training nodes and configurations to demonstrate its scalability on larger and distributed systems using the OpenImages dataset. We evaluate across the in-house, AWS, and Azure servers to cover a wide range of training systems. We test on the in-house and Azure servers to evaluate multi-node distributed training and on the in-house, AWS, and Azure servers to evaluate concurrent training across three distinct training systems.


We evaluate Seneca's performance on distributed systems by measuring single-job training throughput on two in-house and two Azure servers in distributed data parallel mode with remote caching (Figure~\ref{fig:distributed}).
Network bandwidth and gradient communication are key considerations in this setup.
On 2$\times$ in-house servers, Seneca is 1.6$\times$ faster than MINIO and 1.62$\times$ faster than Seneca on a single in-house server. While gradient communication overhead is negligible due to ring-reduce~\cite{tang2020communication}, the 10 Gbps network bottlenecks throughput, limiting scaling to 1.62$\times$ instead of the expected $\approx2\times$. On an Azure server with 80 Gbps, this bottleneck is eliminated, allowing Seneca to scale by 1.89$\times$ going from one to two nodes. On 2$\times$Azure nodes, Seneca outperforms MINIO, the next best by 42.39\%.

\begin{figure}
\centering
        \includegraphics[width=\columnwidth]{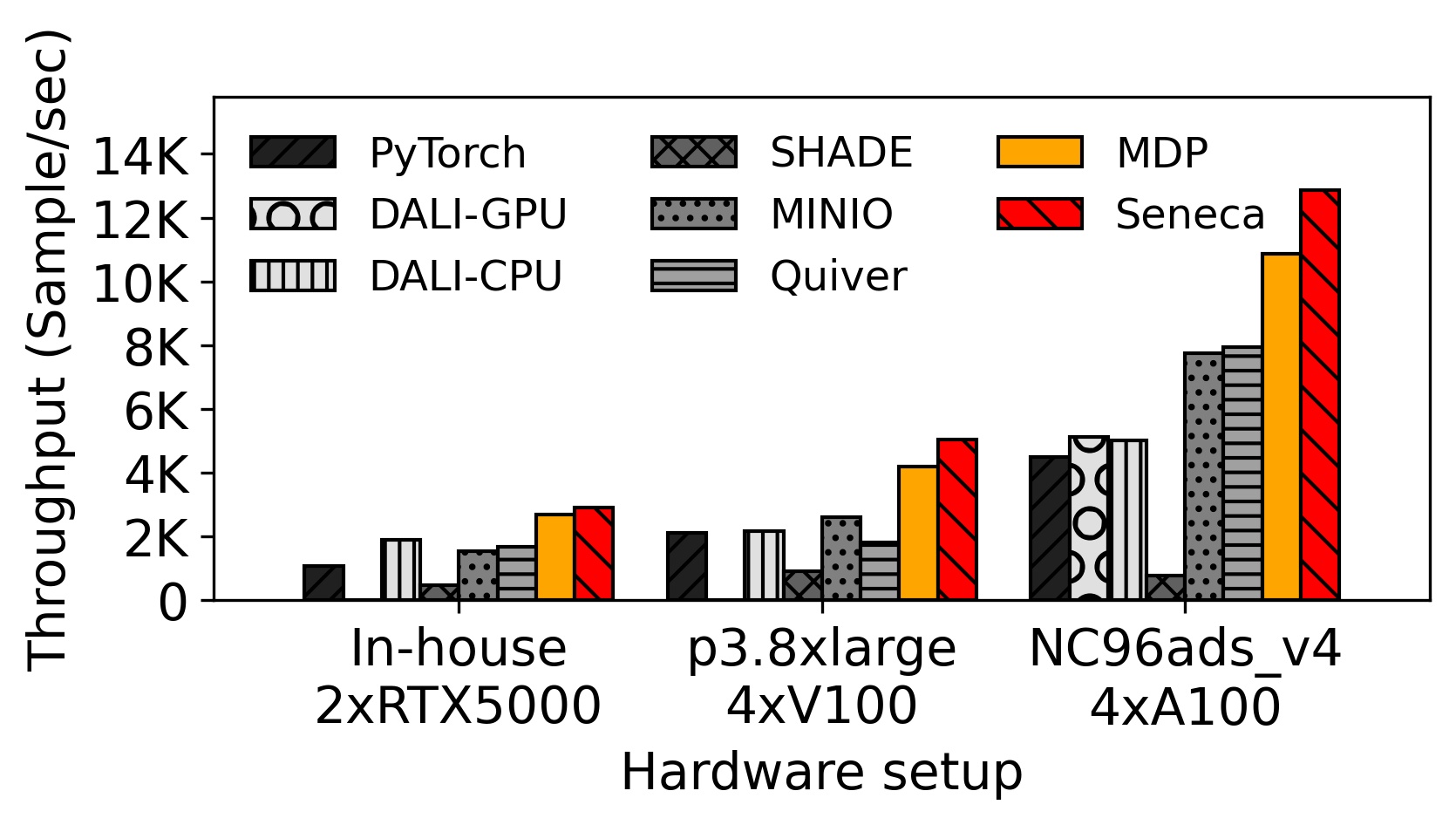}
  \caption{
        Throughput for training 2 jobs concurrently on different hardware platforms. Seneca performs well across a wide range of system configurations.
  }
  \label{fig:scalability}
\end{figure}

Next, we evaluate Seneca's scalability across various ML training systems. 
Figure~\ref{fig:scalability} shows throughput for two jobs training concurrently on the in-house, AWS, and Azure servers. While the AWS and Azure servers have slower remote storage bandwidths, they are faster overall, with more powerful GPUs, which are better suited for ML training~\cite{markidis2018nvidia} as shown in Table~\ref{tab:nominal_vals}.
We make three observations:
(1) Seneca's throughput increases by 4.44$\times$ when moving from the in-house server (2$\times$RTX5000) to the Azure server (4$\times$A100)
(2) Seneca outperforms the next best dataloader by 1.61$\times$ on Azure (vs. Quiver), 1.93$\times$ on AWS (vs. MINIO), and 1.52$\times$ on the in-house server (vs. DALI-CPU).
(3) DALI-GPU fails for two or more concurrent jobs on the in-house and AWS servers due to insufficient GPU memory.

\begin{figure}
\centering
        \includegraphics[width=\columnwidth]{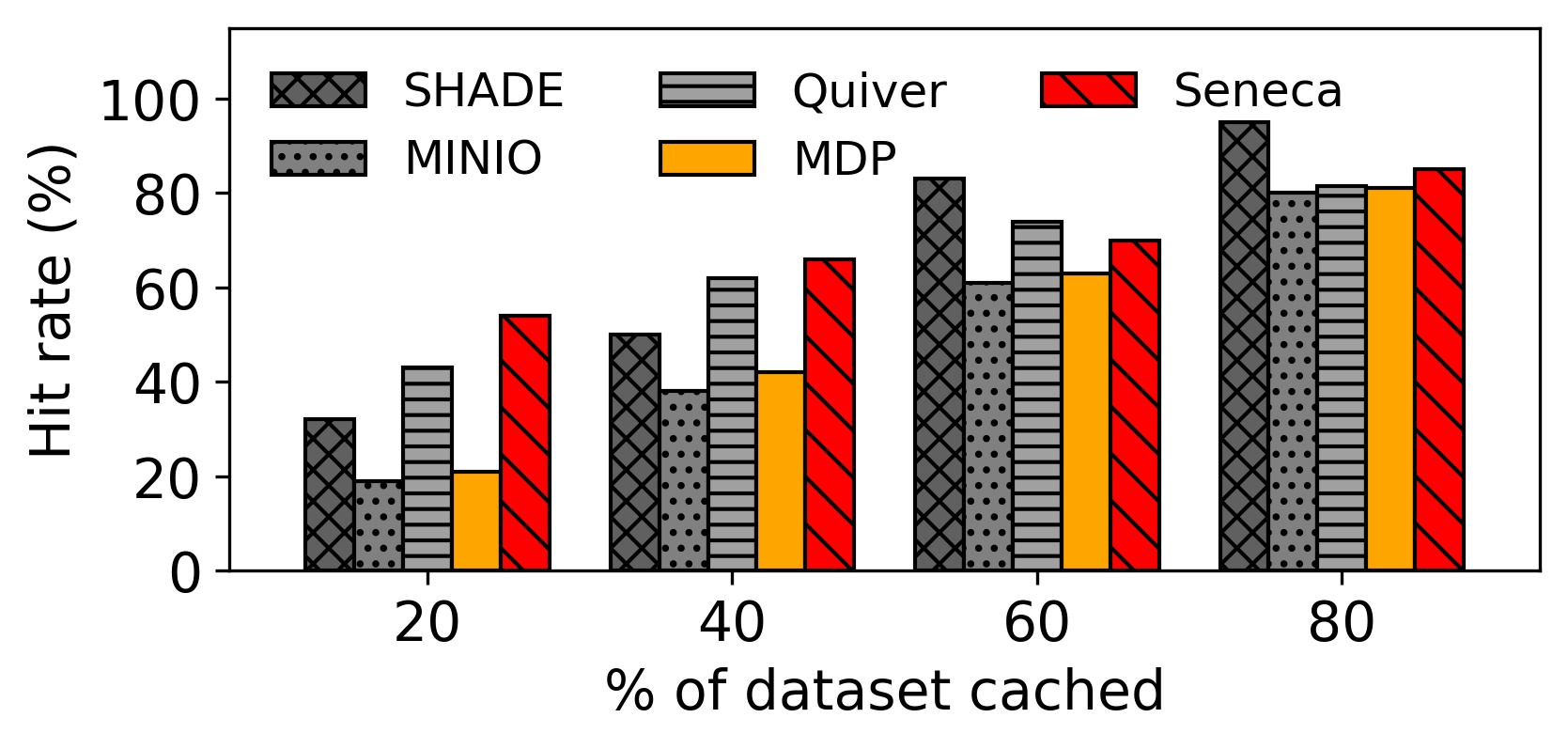}
  \caption{
        Cache hit rate when concurrently training three models. Seneca improves cache efficiency, achieving 54\% hit rate with only 20\% of the dataset cached.
  }
  \label{fig:hitrate_bar}
\end{figure}

Finally, we evaluate Seneca's cache hit rate with varying cache sizes to evaluate its performance with larger datasets and limited cache capacity. 
Figure~\ref{fig:hitrate_bar} presents the cache hit rate, calculated as the total cache hits across all partitions divided by the number of samples in the dataset, during concurrent training of AlexNet, ResNet-50, and MobileNetV2 on ImageNet-1K.
Seneca achieves a 54\% hit rate with just 20\% of the dataset cached, 11\% higher than Quiver, the next best. Seneca improves the hit rate by replacing uncached samples with cached ones not seen in the current epoch. Even with 40\% of the dataset cached, Seneca maintains the highest hit rate of 66\%. With 60\% and 80\% cached, SHADE's hit rate surpasses Seneca but throughput is still the slowest due to it's single threaded design. MINIO and MDP show hit rates roughly equal to the percentage of cached data since they do not implement specialized policies to improve the hit rate.
\subsection{Load sensitivity}\label{sec:sw_scalability}

\begin{figure}
\centering
        \includegraphics[width=\columnwidth]{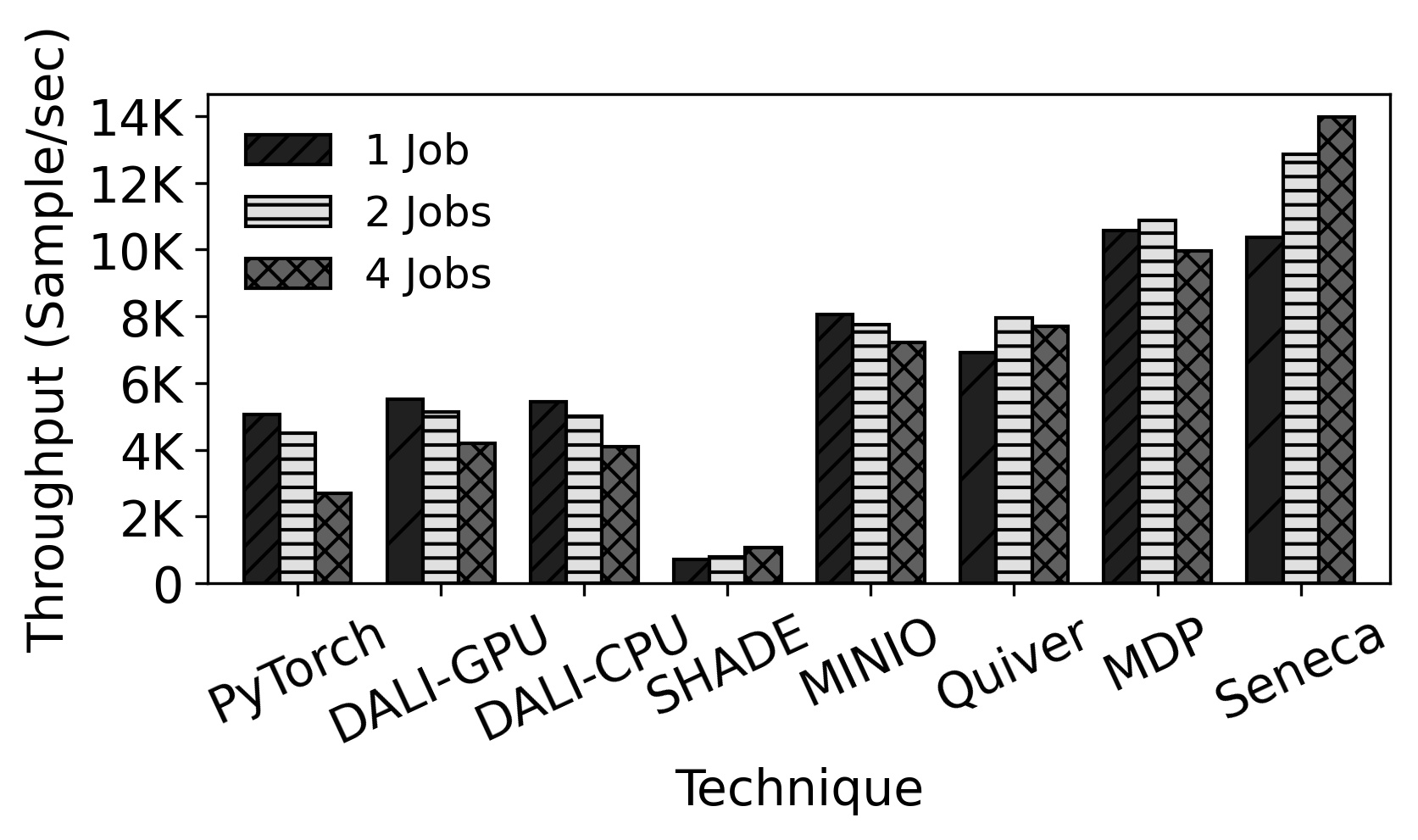}
  \caption{
        Aggregate DSI throughput for up to four concurrent models on the Azure server with 4$\times$A100 GPUs. Seneca boosts DSI pipeline throughput by 1.81$\times$ compared to Quiver for four jobs.
  }
  \label{fig:concurrency}
\end{figure}

\begin{figure*}[t!]
        \begin{subfigure}[b]{0.33\linewidth}
                \includegraphics[width=\columnwidth]{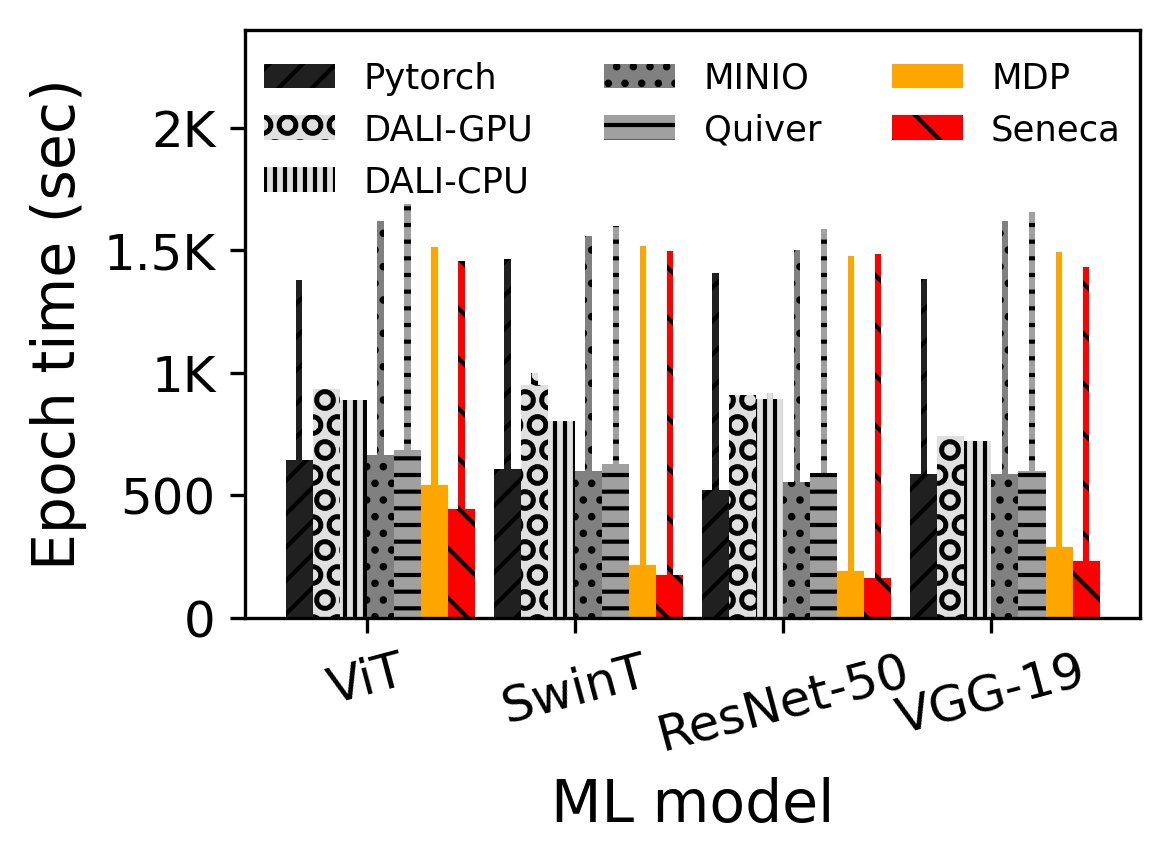}
                \caption{Imagenet-1K on 1$\times$Azure}
                \label{fig:ect_combined_imnet1k}
        \end{subfigure}%
        \hfill
        \begin{subfigure}[b]{0.33\linewidth}
                \includegraphics[width=\columnwidth]{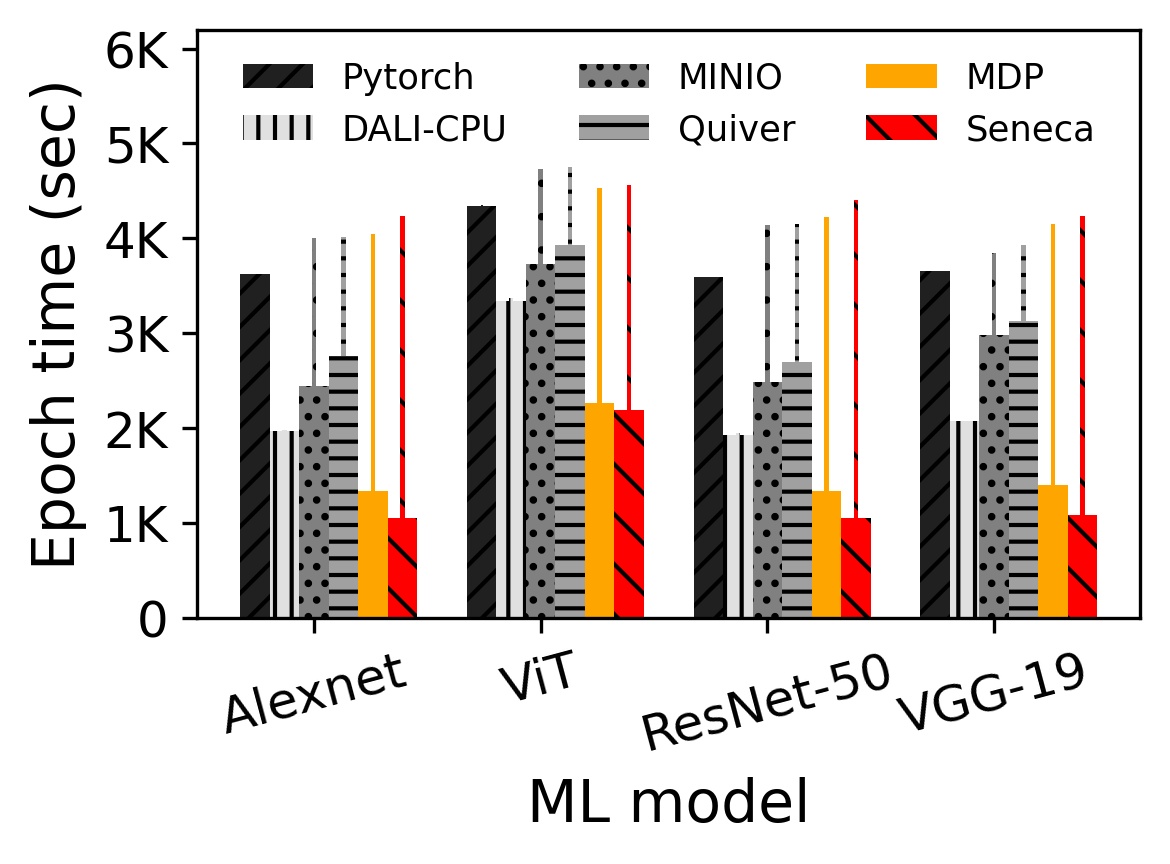}
                \caption{OpenImages on 1$\times$AWS}
                \label{fig:ect_combined_openimages}
        \end{subfigure}
        \hfill
        \begin{subfigure}[b]{0.33\linewidth}
                \includegraphics[width=\columnwidth]{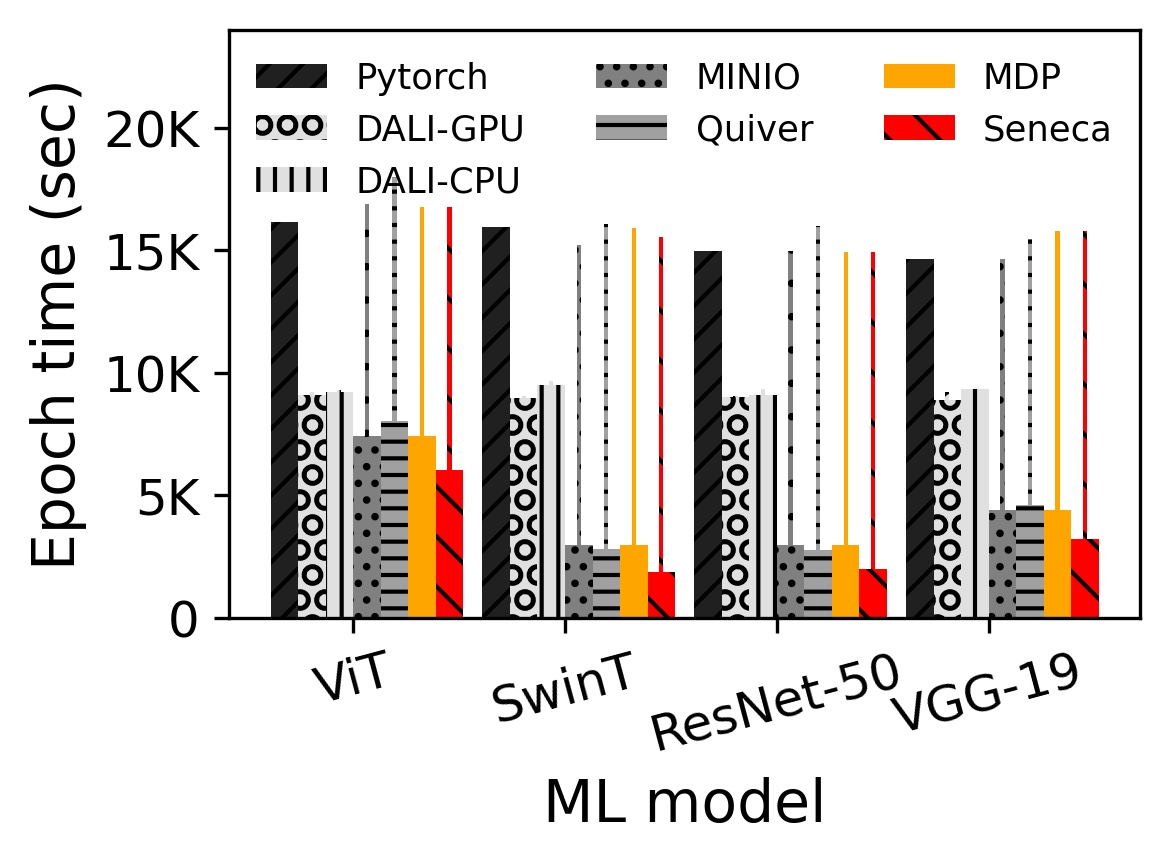}
                \caption{ImageNet-22K on 1$\times$Azure}
                \label{fig:ect_combined_imagenet22k}
        \end{subfigure}
        \caption{
        Stable epoch completion time (bars) and first-epoch time (lines) for concurrent training of popular DNN models (two jobs) across different datasets, servers, and dataloaders. The first epoch reflects training when the cache is not warmed up, while subsequent epochs represent training with the cache in use. Lower is better.
        }\label{fig:ect_combined}
\end{figure*}

\vspace{10pt}
\begin{table}
    \centering
    \footnotesize
    \caption{CPU and GPU utilization for four concurrently training jobs on the in-house server using various dataloaders.
    }
    \begin{tabular*}{\columnwidth}{@{\extracolsep{\fill}}lcccccc@{}}
        \toprule
        
        & \makecell[cc]{PyTorch}
        & \makecell[cc]{DALI-CPU}
        & \makecell[cc]{MINIO}
        & \makecell[cc]{Quiver}
        & \makecell[cc]{MDP}
        & \makecell[cc]{Seneca}\\
        \midrule
        CPU & 88\% & 88\% & 91\% & 96\% & 43\% & \textbf{54\%} \\
        GPU & 72\% & 76\% & 79\% & 80\% & 98\% & \textbf{98\%} \\
        \bottomrule
    \end{tabular*}
    \label{tab:gpu_cpu_util}
\end{table}

We evaluate Seneca's impact on concurrent model training by measuring DSI pipeline throughput on the Azure server with 400 GB remote cache while varying the number of concurrent jobs (Figure~\ref{fig:concurrency}). We use the OpenImages dataset as it is larger than the size of the remote cache and thus represents a common scenario. We also show the GPU and CPU utilization for four concurrently training jobs in Table~\ref{tab:gpu_cpu_util} to show that Seneca saturates GPU resources.

We make five observations: 
(1) Our optimizations, MDP and Seneca outperform all other dataloaders by at least 28.97\% (single job on MINIO vs single job on Seneca).
(2) For four concurrently training jobs, Seneca is the best performing and outperforms Quiver, the next best performing dataloader by 1.81$\times$. 
(3) For four or more concurrent jobs on a single Azure server, the performance of Seneca is bounded by the GPU at 98\% utilization (Table~\ref{tab:gpu_cpu_util}), limiting speedup with added concurrency. 
(4) PyTorch, DALI, MINIO, and Quiver are limited by I/O and CPU preprocessing and do not scale well with concurrent jobs. 
(5) Seneca outperforms SHADE by 13.18$\times$ due to SHADE's single-threaded design.

\subsection{ML model and dataset sensitivity}\label{sec:epoch_time}


Finally, we show Seneca's generality across different ML models, datasets, and hardware setups. For each dataloader, we train two models concurrently and measure the first epoch completion time (ECT) and the average ECT for subsequent epochs. 
The first epoch time represents execution with cold page and user-level caches, while subsequent epoch times reflect execution with warmed-up caches.
We exclude SHADE due to its single-threaded design. For this evaluation, we use two transformer models--ViT~\cite{dosovitskiy2020image}, and SwinT~\cite{liu2021swin} and three DNN models--VGG-19, ResNet-50, and AlexNet to cover a wide range of model architectures.
\vspace{7pt}

We make three observations for training with ImageNet-1K on one Azure server with 400 GB remote cache (Figure~\ref{fig:ect_combined_imnet1k}):
(1) Seneca's stable ECT for the largest vision transformer (ViT-h) is 31.36\% lower than the next best dataloader, PyTorch while for ResNet-50, it is 3.45$\times$ faster than MINIO.
(2) When the dataset size (142 GB) is significantly smaller than DRAM size (880 GB), the entire dataset is in page-cache and PyTorch's stable ECT is faster than DALI by at least 31.36\% (VGG-19).
(3) MINIO and Quiver's encoded data cache cannot mitigate redundant decoding and augmentation on the CPU, reducing the gains from their optimizations.


Next, to understand the impact of larger sample sizes (2.75$\times$ larger than ImageNet-1K) on training, we examine the ECTs for models shown in Figure~\ref{fig:ect_combined_openimages} with the OpenImages dataset on the AWS server with 400 GB remote cache.
We make four observations: 
(1) On the AWS system, DSI pipeline is a bottleneck due to limited I/O bandwidth and CPU threads. On this system, training VGG-19 with Seneca is 1.91$\times$ faster than with DALI-CPU.
(2) By caching more data samples in the decoded form to mitigate the higher DSI overheads, Seneca's stable ECT is lower by 87.22\% for AlexNet, 87.15\% for ViT, 85.53\% for ResNet-50, and 47.85\% for VGG-19 compared to the DALI-CPU, the next best dataloader. 
(3) Caching dataloaders (MINIO, and Quiver) reduce fetch stalls by caching encoded data, reducing stable ECT by up to 39.85\% (ResNet-50 on MINIO).
(4) On the V100 GPUs of the AWS server, DALI-GPU fails for concurrent training jobs due to limited GPU memory.


For ImageNet-22K (1.4 TB) on the Azure server with 400 GB remote cache (Figure~\ref{fig:ect_combined_imagenet22k}), we make four observations:
(1) PyTorch, DALI-CPU, and DALI-GPU which depend on the page-cache perform poorly for large datasets.
(2) Due to the limited cache size, MDP allocates 100\% of the available memory for caching raw data, thus performing similar to MINIO.
(3) Thanks to ODS, Seneca's ECT is 29.35\% lower on average when compared to the next best dataloader across all evaluated models.
(4) Seneca reduces stable ECT for SwinT by 8.37$\times$ by mitigating I/O and preprocessing bottlenecks.



\section{Conclusion}
We present Seneca, a cache partitioning and data sampling system for optimizing concurrent DNN training on distributed systems.
Seneca consists of two techniques: 
model-driven partitioning that finds the best way to partition the cache to improve DSI throughput 
and opportunistic data sampling that preferentially selects cached data samples to maximize the cache hit rate. 
In doing so, Seneca reduces model training time and multi-job makespan by 45.23\% 
and improves the DSI pipeline throughput by up to 3.45$\times$
on concurrent jobs.

\section*{Acknowledgments}

We would like to thank our shepherd, Ali Anwar, and the anonymous reviewers for their constructive feedback. We also thank Syracuse University Information Technology Services (ITS) for providing computing resources and support. This research was supported in part by the National Science Foundation under award numbers CNS-2008453, CSR-2323100, CAREER-2338457, CSR-2402328, and CAREER-2443515, and by Samsung through the Alternative Sustainable and Intelligent Computing Industry-University Cooperative Research Center (NSF \#1822165). This work is also supported through the Global Research Support Program in the Digital Field supervised by the Institute for Information and Communications Technology Planning and Evaluation (IITP) under Grant RS-2024-00428758, funded by the Ministry of Science and ICT (MSIT), South Korea. Any opinions, findings, conclusions, or recommendations expressed in this material are those of the authors and do not necessarily reflect the views of the supporting agencies.

\bibliographystyle{plain}
\bibliography{bibliography}

\begin{thebibliography}{10}

\bibitem{fio}
Flexible {I/O} tester.
\newblock \url{https://github.com/axboe/fio}.

\bibitem{imagenet22k}
{ImageNet-22K}.
\newblock \url{https://image-net.org/download-images.php}.

\bibitem{pytorch}
{PyTorch}.
\newblock \url{https://pytorch.org}.

\bibitem{redis}
Redis.
\newblock \url{https://redis.io}.

\bibitem{tensorflow2015-whitepaper}
Mart\'{i}n Abadi, Ashish Agarwal, Paul Barham, Eugene Brevdo, Zhifeng Chen, Craig Citro, Greg~S. Corrado, Andy Davis, Jeffrey Dean, Matthieu Devin, Sanjay Ghemawat, Ian Goodfellow, Andrew Harp, Geoffrey Irving, Michael Isard, Yangqing Jia, Rafal Jozefowicz, Lukasz Kaiser, Manjunath Kudlur, Josh Levenberg, Dandelion Man\'{e}, Rajat Monga, Sherry Moore, Derek Murray, Chris Olah, Mike Schuster, Jonathon Shlens, Benoit Steiner, Ilya Sutskever, Kunal Talwar, Paul Tucker, Vincent Vanhoucke, Vijay Vasudevan, Fernanda Vi\'{e}gas, Oriol Vinyals, Pete Warden, Martin Wattenberg, Martin Wicke, Yuan Yu, and Xiaoqiang Zheng.
\newblock {TensorFlow}: Large-scale machine learning on heterogeneous systems, 2015.
\newblock Software available from \url{https://www.tensorflow.org/}.

\bibitem{AudibertCGKST23socc}
Andrew Audibert, Yang Chen, Dan Graur, Ana Klimovic, Jir{\'{\i}} Simsa, and Chandramohan~A. Thekkath.
\newblock tf.data service: A case for disaggregating {ML} input data processing.
\newblock In {\em {ACM} Symposium on Cloud Computing (SoCC)}, pages 358--375, 2023.
\newblock \url{https://doi.org/10.1145/3620678.3624666}.

\bibitem{awsp2instancetype}
Amazon Web~Services (AWS).
\newblock Amazon {EC2} {P2} instances. powerful, scalable {GPU} instances for high-performance computing.
\newblock \url{https://aws.amazon.com/ec2/instance-types/p2/}, 2023.

\bibitem{awsp3instancetype}
Amazon Web~Services (AWS).
\newblock Amazon {EC2} {P3} instances. accelerate machine learning and high performance computing applications with powerful {GPUs}.
\newblock \url{https://aws.amazon.com/ec2/instance-types/p3/}, 2023.

\bibitem{awsx2i}
Amazon Web~Services (AWS).
\newblock {AWS} {EC2} {X1} instance types.
\newblock \url{https://aws.amazon.com/ec2/instance-types/x2i/}, 2024.

\bibitem{azureE}
Azure.
\newblock {'E'} family memory optimized {VM} size series.
\newblock \url{https://learn.microsoft.com/en-us/azure/virtual-machines/sizes/memory-optimized/e-family?tabs=epsv6%2Ceasv6%2Cev5%2Cedv5%2Ceasv5%2Cepsv5}, 2025.

\bibitem{AzureNCv4}
Azure.
\newblock {NC_A100_v4} sizes series.
\newblock \url{https://learn.microsoft.com/en-us/azure/virtual-machines/sizes/gpu-accelerated/nca100v4-series?tabs=sizebasic}, 2025.

\bibitem{bao2020preemptive}
Yixin Bao, Yanghua Peng, Yangrui Chen, and Chuan Wu.
\newblock Preemptive all-reduce scheduling for expediting distributed dnn training.
\newblock In {\em IEEE INFOCOM 2020-IEEE Conference on Computer Communications}, pages 626--635. IEEE, 2020.
\newblock \url{https://ieeexplore.ieee.org/document/9155446}.

\bibitem{chen2023icache}
Weijian Chen, Shuibing He, Yaowen Xu, Xuechen Zhang, Siling Yang, Shuang Hu, Xian-He Sun, and Gang Chen.
\newblock {iCache}: An importance-sampling-informed cache for accelerating {I/O}-bound {DNN} model training.
\newblock In {\em 2023 IEEE International Symposium on High-Performance Computer Architecture (HPCA)}, pages 220--232. IEEE, 2023.
\newblock \url{https://ieeexplore.ieee.org/document/10070964}.

\bibitem{dean2012large}
Jeffrey Dean, Greg Corrado, Rajat Monga, Kai Chen, Matthieu Devin, Mark Mao, Marc'aurelio Ranzato, Andrew Senior, Paul Tucker, Ke~Yang, et~al.
\newblock Large scale distributed deep networks.
\newblock {\em Advances in neural information processing systems}, 25, 2012.
\newblock \url{https://dl.acm.org/doi/10.5555/2999134.2999271}.

\bibitem{deng2009imagenet}
Jia Deng, Wei Dong, Richard Socher, Li-Jia Li, Kai Li, and Li~Fei-Fei.
\newblock {ImageNet}: A large-scale hierarchical image database.
\newblock In {\em IEEE Conference on Computer Vision and Pattern Recognition (CVPR)}, pages 248--255, 2009.
\newblock \url{https://doi.org/10.1109/CVPR.2009.5206848}.

\bibitem{dosovitskiy2020image}
Alexey Dosovitskiy.
\newblock An image is worth 16x16 words: Transformers for image recognition at scale.
\newblock {\em arXiv preprint arXiv:2010.11929}, 2020.
\newblock \url{https://arxiv.org/pdf/2010.11929/1000}.

\bibitem{dryden2021clairvoyant}
Nikoli Dryden, Roman B{\"{o}}hringer, Tal Ben{-}Nun, and Torsten Hoefler.
\newblock Clairvoyant prefetching for distributed machine learning {I/O}.
\newblock In {\em ACM International Conference for High Performance Computing, Networking, Storage and Analysis (SC)}, pages 92:1--92--15, 2021.
\newblock \url{https://doi.org/10.1145/3458817.3476181}.

\bibitem{duplyakin2019design}
Dmitry Duplyakin, Robert Ricci, Aleksander Maricq, Gary Wong, Jonathon Duerig, Eric Eide, Leigh Stoller, Mike Hibler, David Johnson, Kirk Webb, et~al.
\newblock The design and operation of {CloudLab}.
\newblock In {\em USENIX Annual Technical Conference (ATC)}, pages 1--14, 2019.
\newblock \url{https://www.usenix.org/conference/atc19/presentation/duplyakin}.

\bibitem{gcp}
GCP.
\newblock {GCP} {GPU} platforms.
\newblock \url{https://cloud.google.com/compute/docs/gpus#nvidia_p100_gpus}, 2024.

\bibitem{graur2022cachew}
Dan Graur, Damien Aymon, Dan Kluser, Tanguy Albrici, Chandramohan~A Thekkath, and Ana Klimovic.
\newblock Cachew: Machine learning input data processing as a service.
\newblock In {\em USENIX Annual Technical Conference (ATC)}, pages 689--706, 2022.
\newblock \url{https://www.usenix.org/conference/atc22/presentation/graur}.

\bibitem{graur2024pecan}
Dan Graur, Oto Mraz, Muyu Li, Sepehr Pourghannad, Chandramohan~A Thekkath, and Ana Klimovic.
\newblock Pecan: Cost-efficient ml data preprocessing with automatic transformation ordering and hybrid placement.
\newblock In {\em USENIX Annual Technical Conference (ATC)}, pages 649--665, 2024.
\newblock \url{https://www.usenix.org/conference/atc24/presentation/graur}.

\bibitem{he2016deep}
Kaiming He, Xiangyu Zhang, Shaoqing Ren, and Jian Sun.
\newblock Deep residual learning for image recognition.
\newblock In {\em IEEE Conference on Computer Vision and Pattern Recognition (CVPR)}, pages 770--778, 2016.
\newblock \url{https://doi.org/10.1109/CVPR.2016.90}.

\bibitem{huang2019gpipe}
Yanping Huang, Youlong Cheng, Ankur Bapna, Orhan Firat, Dehao Chen, Mia Chen, HyoukJoong Lee, Jiquan Ngiam, Quoc~V Le, Yonghui Wu, et~al.
\newblock Gpipe: Efficient training of giant neural networks using pipeline parallelism.
\newblock {\em Advances in neural information processing systems}, 32, 2019.
\newblock \url{https://proceedings.neurips.cc/paper_files/paper/2019/hash/093f65e080a295f8076b1c5722a46aa2-Abstract.html}.

\bibitem{isenko2022my}
Alexander Isenko, Ruben Mayer, Jeffrey Jedele, and Hans-Arno Jacobsen.
\newblock Where is my training bottleneck? hidden trade-offs in deep learning preprocessing pipelines.
\newblock In {\em Proceedings of the 2022 International Conference on Management of Data}, pages 1825--1839, 2022.
\newblock \url{https://dl.acm.org/doi/abs/10.1145/3514221.3517848}.

\bibitem{katharopoulos2018not}
Angelos Katharopoulos and Fran{\c{c}}ois Fleuret.
\newblock Not all samples are created equal: Deep learning with importance sampling.
\newblock In {\em PMLR International Conference on Machine Learning (ICML)}, pages 2525--2534, 2018.
\newblock \url{http://proceedings.mlr.press/v80/katharopoulos18a.html}.

\bibitem{khan2023shade}
Redwan Ibne~Seraj Khan, Ahmad~Hossein Yazdani, Yuqi Fu, Arnab~K Paul, Bo~Ji, Xun Jian, Yue Cheng, and Ali~R Butt.
\newblock {SHADE}: Enable fundamental cacheability for distributed deep learning training.
\newblock In {\em USENIX Conference on File and Storage Technologies (FAST)}, pages 135--152, 2023.
\newblock \url{https://www.usenix.org/conference/fast23/presentation/khan}.

\bibitem{kim2023fusionflow}
Taeyoon Kim, ChanHo Park, Mansur Mukimbekov, Heelim Hong, Minseok Kim, Ze~Jin, Changdae Kim, Ji-Yong Shin, and Myeongjae Jeon.
\newblock Fusionflow: Accelerating data preprocessing for machine learning with {CPU-GPU} cooperation.
\newblock {\em Proceedings of the VLDB Endowment}, 17(4):863--876, 2023.
\newblock \url{https://doi.org/10.14778/3636218.3636238}.

\bibitem{OpenImages}
Ivan Krasin, Tom Duerig, Neil Alldrin, Vittorio Ferrari, Sami Abu-El-Haija, Alina Kuznetsova, Hassan Rom, Jasper Uijlings, Stefan Popov, Siyamalan Kamali, Matteo Malloci, Jordi Pont-Tuset, Andreas Veit, Serge Belongie, Vicente Gomes, Abhinav Gupta, Chen Sun, Gal Chechik, David Cai, Zheyun Feng, Dhyanesh Narayanan, and Kevin Murphy.
\newblock Openimages dataset.
\newblock \url{https://storage.googleapis.com/openimages/web/index.html}, 2017.

\bibitem{krizhevsky2012imagenet}
Alex Krizhevsky, Ilya Sutskever, and Geoffrey~E. Hinton.
\newblock {ImageNet} classification with deep convolutional neural networks.
\newblock In {\em Conference on Neural Information Processing Systems (NIPS)}, pages 1106--1114, 2012.
\newblock \url{https://proceedings.neurips.cc/paper/2012/hash/c399862d3b9d6b76c8436e924a68c45b-Abstract.html}.

\bibitem{kuchnik2022plumber}
Michael Kuchnik, Ana Klimovic, Jiri Simsa, Virginia Smith, and George Amvrosiadis.
\newblock Plumber: Diagnosing and removing performance bottlenecks in machine learning data pipelines.
\newblock {\em Proceedings of Machine Learning and Systems}, 4:33--51, 2022.
\newblock \url{https://proceedings.mlsys.org/paper_files/paper/2022/hash/d0e90e9a9310570dfa643aa3b2da6e89-Abstract.html}.

\bibitem{kumar2020quiver}
Abhishek~Vijaya Kumar and Muthian Sivathanu.
\newblock Quiver: An informed storage cache for deep learning.
\newblock In {\em Conference on File and Storage Technologies (FAST)}, pages 283--296, 2020.
\newblock \url{https://www.usenix.org/conference/fast20/presentation/kumar}.

\bibitem{lecun2002efficient}
Yann LeCun, L{\'e}on Bottou, Genevieve~B Orr, and Klaus-Robert M{\"u}ller.
\newblock Efficient backprop.
\newblock In {\em Neural networks: Tricks of the trade}, pages 9--50. Springer, 2002.

\bibitem{lee2021refurbish}
Gyewon Lee, Irene Lee, Hyeonmin Ha, Kyunggeun Lee, Hwarim Hyun, Ahnjae Shin, and Byung-Gon Chun.
\newblock Refurbish your training data: Reusing partially augmented samples for faster deep neural network training.
\newblock In {\em USENIX Annual Technical Conference (ATC)}, pages 537--550, 2021.
\newblock \url{https://www.usenix.org/conference/atc21/presentation/lee}.

\bibitem{li2014communication}
Mu~Li, David~G Andersen, Alexander~J Smola, and Kai Yu.
\newblock Communication efficient distributed machine learning with the parameter server.
\newblock {\em Advances in Neural Information Processing Systems}, 27, 2014.
\newblock \url{https://papers.nips.cc/paper_files/paper/2014/hash/935ad074f32d1e8f085a143449894cdc-Abstract.html}.

\bibitem{lim2021zico}
Gangmuk Lim, Jeongseob Ahn, Wencong Xiao, Youngjin Kwon, and Myeongjae Jeon.
\newblock Zico: Efficient {GPU} memory sharing for concurrent {DNN} training.
\newblock In {\em 2021 USENIX Annual Technical Conference (USENIX ATC 21)}, pages 161--175, 2021.
\newblock \url{https://www.usenix.org/conference/atc21/presentation/lim}.

\bibitem{liu2022lobster}
Jie Liu, Bogdan Nicolae, and Dong Li.
\newblock Lobster: Load balance-aware {I/O} for distributed {DNN} training.
\newblock In {\em ACM International Conference on Parallel Processing (ICPP)}, pages 1--11, 2022.
\newblock \url{https://dl.acm.org/doi/abs/10.1145/3545008.3545090}.

\bibitem{liu2021swin}
Ze~Liu, Yutong Lin, Yue Cao, Han Hu, Yixuan Wei, Zheng Zhang, Stephen Lin, and Baining Guo.
\newblock Swin transformer: Hierarchical vision transformer using shifted windows.
\newblock In {\em Proceedings of the IEEE/CVF international conference on computer vision}, pages 10012--10022, 2021.
\newblock \url{https://ieeexplore.ieee.org/document/9710580}.

\bibitem{markidis2018nvidia}
Stefano Markidis, Steven~Wei Der~Chien, Erwin Laure, Ivy~Bo Peng, and Jeffrey~S Vetter.
\newblock {Nvidia} tensor core programmability, performance \& precision.
\newblock In {\em 2018 IEEE international parallel and distributed processing symposium workshops (IPDPSW)}, pages 522--531. IEEE, 2018.
\newblock \url{https://ieeexplore.ieee.org/document/8425458}.

\bibitem{dsanalyzer}
Jayashree Mohan.
\newblock {MSR-fiddle/DS-Analyzer}.
\newblock \url{https://github.com/msr-fiddle/DS-Analyzer/tree/main}, 2021.

\bibitem{mohan2022looking}
Jayashree Mohan, Amar Phanishayee, Janardhan Kulkarni, and Vijay Chidambaram.
\newblock Looking beyond {GPUs} for {DNN} scheduling on multi-tenant clusters.
\newblock In {\em USENIX Symposium on Operating Systems Design and Implementation (OSDI)}, pages 579--596, 2022.
\newblock \url{https://www.usenix.org/conference/osdi22/presentation/mohan}.

\bibitem{mohan2020analyzing}
Jayashree Mohan, Amar Phanishayee, Ashish Raniwala, and Vijay Chidambaram.
\newblock Analyzing and mitigating data stalls in {DNN} training.
\newblock {\em Proceedings of the Very Large Data Bases Endowment (PVLDB)}, 14(5):771--784, 2021.
\newblock \url{http://www.vldb.org/pvldb/vol14/p771-mohan.pdf}.

\bibitem{murray2021tf}
Derek~Gordon Murray, Jiri Simsa, Ana Klimovic, and Ihor Indyk.
\newblock tf.data: A machine learning data processing framework.
\newblock {\em Proceedings of Very Large Data Bases Endowment (PVLDB)}, 14(12):2945--2958, 2021.
\newblock \url{http://www.vldb.org/pvldb/vol14/p2945-klimovic.pdf}.

\bibitem{nvidia-dali}
NVIDIA.
\newblock {NVIDIA} {Data} {Loading} {Library}.
\newblock \url{https://developer.nvidia.com/dali}, 2023.

\bibitem{nvidiaa100}
NVIDIA.
\newblock {NVIDIA A100}.
\newblock \url{https://www.nvidia.com/content/dam/en-zz/Solutions/Data-Center/a100/pdf/nvidia-a100-datasheet-nvidia-us-2188504-web.pdf}, 2024.

\bibitem{nvidiah100}
NVIDIA.
\newblock {NVIDIA H100}.
\newblock \url{https://www.nvidia.com/en-us/data-center/h100/}, 2024.

\bibitem{nvidiap100}
NVIDIA.
\newblock {NVIDIA P100}.
\newblock \url{https://images.nvidia.com/content/tesla/pdf/nvidia-tesla-p100-datasheet.pdf}, 2024.

\bibitem{nvidiak20}
NVIDIA.
\newblock {NVIDIA TESLA K20}.
\newblock \url{https://www.nvidia.com/content/PDF/kepler/NV_DS_TeslaK_Family_May_2012_LR.pdf}, 2024.

\bibitem{nvidiak40}
NVIDIA.
\newblock {NVIDIA TESLA K40}.
\newblock \url{https://www.nvidia.com/content/PDF/kepler/nvidia-tesla-k40.pdf}, 2024.

\bibitem{nvidiak80}
NVIDIA.
\newblock {NVIDIA TESLA K80}.
\newblock \url{https://www.nvidia.com/en-gb/data-center/tesla-k80/}, 2024.

\bibitem{nvidiav100}
NVIDIA.
\newblock {NVIDIA V100}.
\newblock \url{https://images.nvidia.com/content/technologies/volta/pdf/volta-v100-datasheet-update-us-1165301-r5.pdf}, 2024.

\bibitem{nvlink}
NVIDIA.
\newblock {NVLink} and {NVLink Switch}.
\newblock \url{https://www.nvidia.com/en-us/data-center/nvlink/}, 2024.

\bibitem{park2020trainbox}
Pyeongsu Park, Heetaek Jeong, and Jangwoo Kim.
\newblock Trainbox: an extreme-scale neural network training server architecture by systematically balancing operations.
\newblock In {\em 2020 53rd Annual IEEE/ACM International Symposium on Microarchitecture (MICRO)}, pages 825--838. IEEE, 2020.
\newblock \url{https://ieeexplore.ieee.org/document/9251955}.

\bibitem{paszke2019pytorch}
Adam Paszke, Sam Gross, Francisco Massa, Adam Lerer, James Bradbury, Gregory Chanan, Trevor Killeen, Zeming Lin, Natalia Gimelshein, Luca Antiga, et~al.
\newblock Pytorch: An imperative style, high-performance deep learning library.
\newblock {\em Advances in neural information processing systems}, 32, 2019.
\newblock \url{https://proceedings.neurips.cc/paper/2019/hash/bdbca288fee7f92f2bfa9f7012727740-Abstract.html}.

\bibitem{pinto2018hoard}
Christian Pinto, Yiannis Gkoufas, Andrea Reale, Seetharami Seelam, and Steven Eliuk.
\newblock Hoard: A distributed data caching system to accelerate deep learning training on the cloud.
\newblock {\em arXiv preprint arXiv:1812.00669}, 2018.
\newblock \url{https://doi.org/10.48550/arXiv.1812.00669}.

\bibitem{raina2009large}
Rajat Raina, Anand Madhavan, and Andrew~Y Ng.
\newblock Large-scale deep unsupervised learning using graphics processors.
\newblock In {\em Proceedings of the 26th annual international conference on machine learning}, pages 873--880, 2009.
\newblock \url{https://dl.acm.org/doi/10.1145/1553374.1553486}.

\bibitem{tang2020communication}
Zhenheng Tang, Shaohuai Shi, Wei Wang, Bo~Li, and Xiaowen Chu.
\newblock Communication-efficient distributed deep learning: A comprehensive survey.
\newblock {\em arXiv preprint arXiv:2003.06307}, 2020.
\newblock \url{https://doi.org/10.48550/arXiv.2003.06307}.

\bibitem{um2023fastflow}
Taegeon Um, Byungsoo Oh, Byeongchan Seo, Minhyeok Kweun, Goeun Kim, and Woo-Yeon Lee.
\newblock {FastFlow}: Accelerating deep learning model training with smart offloading of input data pipeline.
\newblock {\em Proceedings of the Very Large Data Bases Endowment (PVLDB)}, 16(5):1086--1099, 2023.
\newblock \url{https://www.vldb.org/pvldb/vol16/p1086-um.pdf}.

\bibitem{weng2022mlaas}
Qizhen Weng, Wencong Xiao, Yinghao Yu, Wei Wang, Cheng Wang, Jian He, Yong Li, Liping Zhang, Wei Lin, and Yu~Ding.
\newblock {MLaaS} in the wild: Workload analysis and scheduling in large-scale heterogeneous {GPU} clusters.
\newblock In {\em 19th USENIX Symposium on Networked Systems Design and Implementation (NSDI 22)}, pages 945--960, 2022.
\newblock \url{https://www.usenix.org/conference/nsdi22/presentation/weng}.

\bibitem{zhao2023goldminer}
Hanyu Zhao, Zhi Yang, Yu~Cheng, Chao Tian, Shiru Ren, Wencong Xiao, Man Yuan, Langshi Chen, Kaibo Liu, Yang Zhang, et~al.
\newblock Goldminer: Elastic scaling of training data pre-processing pipelines for deep learning.
\newblock {\em Proceedings of the ACM on Management of Data}, 1(2):1--25, 2023.
\newblock \url{https://doi.org/10.1145/3589773}.

\bibitem{zhao2024cedar}
Mark Zhao, Emanuel Adamiak, and Christos Kozyrakis.
\newblock cedar: Optimized and unified machine learning input data pipelines.
\newblock {\em Proc. VLDB Endow.}, 2025.
\newblock \url{https://doi.org/10.14778/3705829.3705861}.

\bibitem{zhao2022understanding}
Mark Zhao, Niket Agarwal, Aarti Basant, Bugra Gedik, Satadru Pan, Mustafa Ozdal, Rakesh Komuravelli, Jerry Pan, Tianshu Bao, Haowei Lu, Sundaram Narayanan, Jack Langman, Kevin Wilfong, Harsha Rastogi, Carole{-}Jean Wu, Christos Kozyrakis, and Parik Pol.
\newblock Understanding data storage and ingestion for large-scale deep recommendation model training: industrial product.
\newblock In {\em ACM International Symposium on Computer Architecture (ISCA)}, pages 1042--1057, 2022.
\newblock \url{https://doi.org/10.1145/3470496.3533044}.

\end{thebibliography}

\appendix
\section{Artifact Appendix}

\subsection*{Abstract}

This paper proposes Seneca, a system that alleviates input data preprocessing bottlenecks and reduces training time for concurrent jobs. It introduces two key techniques: (1) a performance model for the data pipeline that optimally partitions the cache among three types of data, and (2) a cache-aware sampling strategy where Seneca prioritizes serving cached data over uncached data during random batch sampling. This artifact includes the code and the steps to train a model using Seneca and comparing it against popular baselines. The experiments have been performed on an in-house server as well as VMs from AWS and Azure.

\subsection*{Scope}
The provided code and scripts support the reproduction and evaluation of the following:

\begin{itemize}
\item Training performance of image models using Seneca under varying cache allocations.
\item Impact of Seneca on final model accuracy.
\item Comparative performance of popular open-source dataloaders, including PyTorch and NVIDIA DALI.
\item Performance evaluation of academic systems such as SHADE, MINIO, and Quiver.
\item Support for training and evaluating multiple models and model types using Seneca.
\end{itemize}

\subsection*{Contents}
\subsubsection{PyTorch and Torchvision}

The implementation accompanying this work is developed using PyTorch, a widely adopted open-source library for machine learning model training and inference, and TorchVision, a companion library that provides efficient image and video data loading and preprocessing utilities.

While the artifact is built on top of PyTorch and TorchVision, the underlying concepts and design are generalizable and can be applied to a wide range of machine learning libraries which have compute-intensive data preprocessing pipelines.

\subsubsection{Redis}

This work uses Redis as the in-memory datastore for caching data samples in various stages of processing—such as encoded, decoded, or pre-processed forms. Redis was chosen for its simplicity, low-latency performance, and broad community support. However, the design is not tied to Redis; any high-performance in-memory key-value store can be used as a drop-in replacement.

The maximum memory allocated for caching can be configured via a flag passed to the training script or by directly modifying the redis configuration file. If the cache is deployed on the same node as the training job, care must be taken to provision sufficient memory to accommodate both the cache and the training process. Alternatively, the cache can be hosted on a separate node or distributed across a cluster of caching nodes, depending on workload size and memory constraints.

\subsubsection{Dataloader comparisons}
Seneca has been evaluated against several dataloaders, including:
\begin{itemize}
    \item PyTorch’s native dataloader
    \item NVIDIA DALI, both with and without GPU offload
    \item Dataloaders from prior academic and industry works, such as SHADE, MINIO, and Quiver
\end{itemize}

All baseline implementations are integrated on top of a common PyTorch version to ensure consistency. Each dataloader can be selected and evaluated using appropriate flags in the provided training script, enabling reproducible and fair comparisons across different data loading backends.

\subsubsection{Docker container}

To simplify setup, we provide a pre-built open-source Docker container with all the necessary tools, dependencies, and libraries required to run Seneca. This container is based on the NVIDIA PyTorch base image from the NGC catalog (\url{https://catalog.ngc.nvidia.com/orgs/nvidia/containers/pytorch}), ensuring compatibility with GPU-accelerated workloads.

You can access the container via the link provided in ~\S~\ref{sec:appendix_hosting}. For step-by-step setup instructions, please refer to the detailed documentation available in our GitHub repository:
\url{https://github.com/swiftomkar/seneca-fast26-pytorch/blob/master/AEC_readme.md}

\subsection*{Hosting}\label{sec:appendix_hosting}

The artifact is available on github repositories:
\begin{itemize}
    \item \textbf{PyTorch:} \url{https://github.com/swiftomkar/seneca-fast26-pytorch}
    \item \textbf{TorchVision:} \url{https://github.com/swiftomkar/seneca-fast26-torchvision}
    \item \textbf{Seneca Docker:} \url{https://hub.docker.com/repository/docker/omkarbdesai/seneca_cuda11.7_cudnn8.5/general}
\end{itemize}

\subsection*{Requirements}
To run Seneca, you will need an NVIDIA GPU with CUDA support. The system has been tested on NVIDIA Quadro RTX 5000, V100, and A100 GPUs. Please ensure the machine has sufficient DRAM for local caching in addition to the memory required by the training process. Alternatively, you may configure a remote caching node.

Our evaluations were conducted on machines with a minimum of 115 GB of DRAM and a 16-core CPU. If you opt for a remote caching setup, ensure that the system has sufficient network bandwidth to maintain optimal performance.

\end{document}